\documentclass[letterpaper,11pt]{JHEP3}

\usepackage{array}
\usepackage{amsmath}
\usepackage{amssymb}
\usepackage{longtable}

\title{Options for Orbifold-GUT Model Building from Five-Dimensional Supergravity}
\author{Sean McReynolds\footnote{sean@phys.psu.edu}\\Physics Department\\ Pennsylvania State University\\ University Park, PA. 16802, USA}

\abstract{This is the first paper of a series that will examine the options for embedding supersymmetric orbifold-GUTs into five-dimensional $\mathcal{N}$=$2$ Yang-Mills-Einstein supergravity theories (YMESGTs).  In particular, we focus on the allowed couplings of charged hypermultiplets in the lowest dimensional reps of the gauge groups $SU(5)$, $SO(10)$ and $E_{6}$.  Our results are within the classification of \textit{homogeneous} quaternionic scalar manifolds.  In the minimal coupling of a generation of bulk matter hypermultiplets, supergravity requires the field content of an $SO(10)$ scenario.  In the minimal coupling of $n$ bulk generations of matter and higgs hypermultiplets, supergravity requires the field content of an $E_{6}$ scenario. 

We also discuss the coupling of tensors and non-compact gaugings in $5D$ YMESGTs, which can serve as alternative ways to obtain four-dimensional Higgs sectors.  Charged tensor couplings seem to be difficult to work with phenomenologically since a $U(1)$ gauge factor is always required when they are present, and it is not clear if tensors can be put in unified multiplets with other fields, if this is desired.  This seems to imply that tensor coulpings in GUT scenarios may be better suited in higher dimensional settings.  The non-compact gaugings discussed here are those of~\cite{GST:85, GZ:03apr}, and offer a novel unification scenario in which the supergravity and vector multiplets are connected by gauge transformations.  

The main points are summarized in tables and the conclusion.  Although the discussion is in the spirit of a ``bottom-up" approach, M-theory is taken as a motivating background.}   

\keywords{GUT, unified theories, supergravity, orbifold}

\preprint{\hepph{0501091}}

\begin{document}

\section{Introduction}

``Grand Unified Theories" (GUTs) seek to explain some of the arbitrary features of the Standard Model, and as a by-product, introduce additional issues that must be resolved.  The proton decay rate is one of these issues, which arises from the new interactions present with unified groups.  Also, a larger unified gauge group requires the introduction of additional massless Higgs fields that are not present in the Standard Model.  Furthermore, assuming unification of the gauge couplings, the value of the weak mixing angle via $\sin^{2}\theta_{W}$ can be predicted and compared with experiment; although suggestively close, the difference is larger than experimental and theoretical uncertainties.

The introduction of rigid spacetime supersymmetry is theoretically and phenomenologically compelling.  It offers a mechanism for stabilization of the electroweak-Planck scale hierarchy (really, the problem becomes: how is supersymmetry spontaneously broken).  As for $4D$ unified theories, the presence of supersymmetry modifies the prediction for $\sin^{2}\theta_{W}$ such that it is in better agreement with the experimental value~\cite{sin2:91}.  The gauge couplings in unified theories meet at a scale that is reasonably close (on a logarithmic scale) to the scale at which gravitational quantum effects are expected to become non-negligible; even closer in the presence of supersymmetry~\cite{sin2:91}.  However, the gravitational and gauge couplings do miss each other in this framework.  But the prediction of the unification scale is an extrapolation based on a $4D$ Yang-Mills field theory.  Therefore, it is at least just as reasonable to propose that the four couplings unify closer to the Planck scale if a quantum theory with gravity is used as the more fundamental theory.  From this point of view, the fact that the gauge couplings and gravitational couplings do not exactly meet in $4D$ susy GUTs indicates that we should be considering corrections due to this more fundamental theory.  For example, heterotic string theory compactified to four dimensions yields a theory whose correction terms can bring the grand unification scale even closer to the characteristic string scale $M_{s}$, and is in fact sufficient to obtain complete coupling unification (an extensive review is~\cite{KD:96feb}, while later discussion can be found in ~\cite{NS:97feb}).  There are downsides to such string unification scenarios, though: Newton's constant is generically unphysically large; and string corrections can change the prediction for $\sin^{2}\theta_{W}$ such that it no longer agrees with experiment.\footnote{It is possible, though, to obtain a prediction for $\sin^{2}\theta_{W}$ that is in good agreement with experiment; see e.g.~\cite{NS:95oct}.}   

The work of Horava and Witten~\cite{HW} offered a possible resolution to these issues.  It was shown that the strong coupling limit of the $10D$ heterotic string theory with $E_{8}\times E_{8}$ gauge group has a description as a weakly coupled $11D$ theory (M-theory) on $M^{10}\times S^{1}/\mathbb{Z}_{2}$.  Further compactification on a Calabi-Yau 3-fold $Y$ yields a theory on $\mathcal{M}^{4}\times S^{1}/\mathbb{Z}_{2}$ with gauge fields that have support only on the boundaries.  Due to the ground state product structure of the spacetime, the size of the spaces can be adjusted independently; setting the size of $S^{1}/\mathbb{Z}_{2}$ to be much larger than $Y$, we obtain an effective $5D$ theory at some intermediate energy scale.  As a result, the running of the gravitational coupling starting from $4D$ is pushed up at the compactification scale $M_{c}$ (the scale of $S^{1}/\mathbb{Z}_{2}$), and all four couplings can meet in the newly unveiled $5D$ theory.  The running of the gauge couplings is unaffected by the fifth dimension since the gauge fields are \textit{confined} to the $4D$ boundary.  This allows a complete unification of couplings in the framework of string theory, without ruining the original predictions of $\sin^{2}\theta_{W}$.  Such scenarios predict a $4D$ Newtonian constant that can have a physically correct order of magnitude, in contrast to compactifications of weakly coupled heterotic string theory.   

These results inspired bottom-up (phenomenological) approaches to dealing with some of the problems in GUTs.  In a number of papers, it was proposed that the the four-dimensional Standard Model is the low energy effective theory of a higher dimensional supersymmetric GUT field theory.  In contrast to earlier Kaluza-Klein unification scenarios (see e.g.~\cite{KK}), these low energy effective theories arise from a theory in a five-dimensional spacetime with boundary; for example, $M^{5}/\mathbb{Z}_{2}$ (for some early papers on this, see~\cite{guts, DM:01aug, HN:01mar}).  In the literature, however, the simplest models have ground state spacetimes isomorphic to $M^{4}\times S^{1}/\mathbb{Z}_{2}$, as in compactifications of the Horava-Witten theory.  Such orbifold-GUTs can suppress proton decay by first eliminating dimension five operators responsible for decay that is too fast; and second by giving large masses to the new fields responsible for the decay.  Additionally, unwanted  scalars in Higgs multiplets of GUT theories can receive large masses, leaving only massless weak doublets.  The presence of boundaries can also perform some or all of the breaking of the GUT group to the Standard Model gauge group.  However, since the gauge fields now live in $5D$, the presence of the fifth dimension affects the running of the gauge couplings, in contrast to the HW scenario.  Therefore, we might worry about the prediction of $\sin^{2}\theta_{W}$~\cite{DDG:98}.       

In light of the above discussion regarding the role of gravitation, the next step from $5D$ supersymmetric orbifold-GUTs is to embed them into $5D$ $\mathcal{N}=2$ supergravity.  General features of $5D$ supergravity orbifolds (without reference to string/M-theory) have been considered in the literature, with couplings to vector and hypermultiplets~\cite{YL:03dec, ZGAZ:04jul, DGKL:04feb}.  However, the question remains, can one realize the field content in orbifold-\textit{GUTs} within the framework of supergravity, and what are the specific theories that do so.  This is the first step in a series of papers that will analyze YMESGTs on orbifold spacetimes, with the motivation being GUT scenarios (from a bottom-up point of view) and strongly coupled string theory (from a top-down point of view).   

Embedding $5D$ supersymmetric GUTs into supergravity is not a trivial incorporation, as $\mathcal{N}=2$ supergravity places restrictive relationships between gaugings and matter content.  It helps to review the situation in the case of \textit{four-dimensional} theories.  In the case of rigidly supersymmetric theories, the set of allowed spin-1/2 multiplet couplings is in one-to-one correspondence with all K\"{a}hler manifolds in the case of $\mathcal{N}=1$ supersymmetry; and hyper-K\"{a}hler manifolds in the case of $\mathcal{N}=2$ supersymmetry.  Local supersymmetry (supergravity) imposes additional restrictions such that the set of allowed spin-1/2 couplings is in one-to-one correspondence with Hodge manifolds, which are special cases of K\"{a}hler manifolds, in the case of $\mathcal{N}=1$ supersymmetry; and one-to-one correspondence with quaternionic manifolds in the case of $\mathcal{N}=2$ supersymmetry.  Thus, in the case of $\mathcal{N}=1$ supersymmetry, only a subset of the possible matter couplings with rigid supersymetry may be directly coupled to supergravity; and in the case of $\mathcal{N}=2$ supersymmetry, none of the possible matter couplings with rigid supersymmetry may be directly coupled to supergravity.  The latter result follows from the fact that quaternionic manifolds have non-zero curvature, while hyper-K\"{a}hler manifolds are flat~\cite{BW:83}.  

Despite the inability to directly couple $\mathcal{N}=2$ supersymmetric theories to supergravity, we can obtain one theory from the other using a particular mapping.  The mapping is between quaternionic and hyper-K\"{a}hler manifolds as the curvature is taken to zero, corresponding to the decoupling of supergravity (the ``rigid limit").  See~\cite{rigid} for the discussion of rigid limits in $4D$ $\mathcal{N}=2$ supergravity.  It has not been proven that a rigid limit always exists.  In fact, it has been shown that there is no rigid limit for supergravity coupled to particular compact scalar manifolds.  If the rigid limit exists, we may take a supergravity theory with compact gauge group coupled to a particular quaternionic scalar manifold, and obtain a particular super-Yang-Mills theory coupled to a hyper-K\"{a}hler scalar manifold.  Though the nature of the hypermultiplet couplings change in this limit, the gauge group representations assigned to the hypermultiplets remain unchanged.

The layout of the paper is as follows.  In section 2, we briefly describe orbifold-GUTs for super Yang-Mills theories.  In section 3 (and in the appendices), we review the framework of $5D$ $\mathcal{N}=2$ supergravity, and provide a list of hypermultiplet couplings for $E_{6}$, $SO(10)$ and $SU(5)$ YMESGTs.  In section 4, we describe an alternative way of obtaining scalars in non-trivial representations of almost-unified gauge groups (necessarily with an abelian factor): by coupling charged tensor multiplets.  In section 5, we comment on non-compact gaugings in supergravity and their possible role in GUTs.  In section 6, we make concluding remarks, summarizing the models that can correspond to the orbifold-GUTs popularly considered in the literature.  Finally, section 7 contains some future directions in which we are working.    

\section{(Rigidly) Supersymmetric Orbifold-GUTs}

In susy orbifold-GUTs, some of the weaknesses of $4D$ unification models are resolved (as briefly discussed in the introducion), while the strengths are left alone where possible.  In particular, the observed fermions already fit nicely into $SU(5)$ representations; or after addition of a gauge singlet fermion, into $SO(10)$.  Even further, the Standard Model Higgs fields can join the fermions (with two additional singlet fermions) to form an irrep of $E_{6}$.  Also, the prediction of $\sin^{2}\theta_{W}$ is not bad in four dimensional susy-GUTs.  Thus, supermultiplets containing the quarks and leptons were originally often given support solely at the four-dimensional fixed planes of the orbifold action.  However, later models allowed at least one generation to lie in the bulk~\cite{YN:01aug, KR:02dec}.  On the other hand, the Higgs sector is not as attractive in four-dimensional GUT theories, as extra massless fields must be introduced to form irreps of the GUT group and large dimension representations are required to spontaneously break the unified group down to the Standard Model. 
By putting the supermultiplets containing the Standard Model Higgs scalars in the five-dimensional bulk, orbifold scenarios can perform part of the breaking, removing the need for a large Higgs sector as well as giving undesirable multiplets large masses.  In contrast to Horava-Witten or Randall-Sundrum type scenarios, orbifold-GUTs have non-abelian gauge multiplets in the bulk giving rise to non-abelian gauge multiplets on the orbifold fixed planes.  This means that the running of gauge couplings will be modified at the scale of compactification for the fifth dimension.  However, in $5D$ orbifold theories, the fifth dimension is the only extra dimension that is ``large", which means the modifications may not be as drastic (this can be seen in the calculations of gauge coupling running throughout the orbifold-GUT literature).  

The $4D$ minimal supersymmetric standard model requires a minimum Higgs supermultiplet content of two chiral multiplets forming the $\mathbf{2}\oplus\mathbf{\bar{2}}$ of $SU(2)$, along with their CPT conjugate supermultiplets.  In fact, this minimum number is preferred by predictions of $\sin^{2}\theta_{W}$~\cite{2higgs}.  However, we do not need to assume such minimality in general; in fact, some constructions outside the framework of orbifold-GUTs prefer non-minimal Higgs coupling~\cite{bailin}.  A $5D$ hypermultiplet $\mathcal{H}$ consists of four scalars and two spin-$1/2$ fields, which would form a pair of $4D$ $\mathcal{N}=1$ chiral multiplets $\{H,H^{c}\}$ and their CPT conjugates.  However, orbifold parity assignments restrict the boundary propagating modes to be the $H$ or $H^{c}$ chiral multiplets.  Therefore, in $SU(5)$ orbifold models the Higgs scalars can be minimally taken to come from $5D$ hypermultiplets in the $\mathbf{5}\oplus\mathbf{5}$ of $SU(5)$.  In $SO(10)$ orbifold models, the scalars are taken to sit in a $5D$ hypermultiplet in the $\mathbf{10}$ of $SO(10)$.  Finally, in $E_{6}$ models, the $5D$ hypermultiplets are taken in the $\mathbf{27}$ of $E_{6}$.  

If these theories are to be low energy effective descriptions of superstring theory, there is a technical point to keep in mind.  To break a GUT group to the Standard Model gauge group solely via a Higgs mechanism requires fields in the adjoint, or larger, representation of the GUT group (as in the model in~\cite{NSW:01apr} described below).  Such a state does not exist below the string scale in (weakly coupled) heterotic string theory if the gauge group is based on a level one Kac-Moody algebra defined on the worldsheet~\cite{DL:89aug, FIQ:90mar}.  Thus, breaking the unified gauge group with a Higgs mechanism at a $4D$ boundary requires a string theory whose $4D$ gauge group is based on a higher level underlying Kac-Moody algebra.  This is not as economical since a large number of additional states, which appear unnecessary, are introduced.\footnote{There are alternative ways to obtain adjoint reps without extra massless matter~\cite{DF:95aug} but these models seem to require a complicated setup.}  In contrast, one can spontaneously break a \textit{partially} unified gauge group\footnote{\textit{Partially} unified gauge group refers to a non-simple group; e.g. $SU(5)\times U(1)$, $SU(4)\times SU(2)\times SU(2)$, or $SU(3)\times SU(3)\times SU(3)$} with a level one spectrum, so all that would remain is to explicitly break the unified group to such a partially unified gauge group using boundary conditions.  Such a mixed scenario has been considered in~\cite{KR:02dec}.  This is one of the benefits of orbifold-GUT scenarios (from the point of view of string theory).

\section{5D YMESGTs and hypermultiplets}

Our goal is to see how the bulk part of orbifold-GUTs can fit into $\mathcal{N}=2$ $5D$ supergravity.  In this section we briefly review the latter's framework (see appendices A,B,C or the references therein for more details).  

\subsection{MESGTs and YMESGTs} 
 
Consider a $5D$ $\mathcal{N}=2$ Maxwell-Einstein supergravity theory (MESGT) with $n_{V}$ vector multiplets~\cite{GST:84}.  The scalars in the vector multiplets parametrize a ``very special" $n_{V}$-dimensional real Riemannian scalar manifold $\mathcal{M}_{R}$, with (possibly trivial) isometry group $Iso(\mathcal{M}_{R})$.  The global symmetry group of the Lagrangian is $G\times SU(2)_{R}$; $G\subset Iso(\mathcal{M}_{R})$ consists of the symmetries of the scalar manifold that are symmetries of the Lagrangian; and $SU(2)_{R}$ is the automorphism group of the $\mathcal{N}=2$ superalgebra.    
 
Consider now the gauging of an $m$-dimensional subgroup $K \subset G \subset  Iso(\mathcal{M}_{R})$, promoting $m$ of the $n_{V}+1$ vector fields of the theory to gauge fields~\cite{GST:85, EGZ:01aug}.  Such theories are called Yang-Mills-Einstein supergravity theories (YMESGTs).  \textit{Note}:  We are not considering the gauging of R-symmetries in this paper; see appendix C for terminology.  Since we are interested only in compact gaugings here, $K$ only has to be a subgroup of the maximal compact subgroup of $G$.  The $n_{V}+1$ vector fields in the (\textit{not necessarily irreducible}) $\mathbf{n_{V}+1}$ of $G$ form representations of $K$ according to the $K\subset G$ decomposition 
\begin{equation}
\mathbf{n+1}\rightarrow \mathbf{adj(K)}\oplus \mathbf{non}\mbox{-}\mathbf{singlets(K)}\oplus \mathbf{singlets(K)}.
\end{equation}
The $m$ vector fields forming the adjoint representation of $K$ can then become the gauge fields of the theory, while all other vector fields in non-trivial reps must be dualized to tensor fields, which satisfy a field equation taking the form of a ``self-duality" constraint (see~\cite{GRW:86, TPvN:84, GZ:99dec} and appendix C.2).  The remaining vector fields, transforming trivially under $K$-transformations, are called ``spectator" vector fields.

Given a particular scalar manifold, one can classify the allowed gaugings and resulting field content via the above analysis.  On the other hand, if a particular gauge group and field content is desired, one can classify the scalar manifolds that allow them.  Before analyzing the problem with particular scalar manifolds in mind, we can state some general features regarding gauging.   These results (and a discussion of them) can be found in~\cite{EGZ:01aug}. 

\begin{itemize}
\item
Clearly, to gauge a compact group $K$, there must be at least $n=\mbox{dim}(K)$ vector supermultiplets.
\item
Gauging a compact semi-simple group $K$ yields at least one $K$-singlet vector field (which can be identified as the graviphoton).  If there is an abelian factor in $K$, the graviphoton can be the corresponding gauge field.
\item
If $K$ is compact and there are non-singlets appearing in eqn (3.1), the corresponding vector fields must be Hodge dualized to tensor fields satisfying ``self-duality" relations.
\item
There can be tensor fields transforming in non-singlet reps of the compact gauge group $K$ iff at least one abelian isometry from $\mathcal{M}_{R}$ is gauged.\footnote{The tensor fields must then at least be charged under the abelian factor, but can transform non-trivially under the other factors as well, according to the decomposition in (3.1).} 

\end{itemize}

In the appendix A, we review the fact that a rank-3 symmetric tensor $C_{IJK}$ defines a pure MESGT completely.
From the discussion in appendix C.1, it is clear that compact gauge groups with singlet graviphoton may be obtained, with no other vector fields in singlet or other non-singlet $K$-reps.  However, we should note that there can be important implications in four dimensions (either in the dimensional reduction or orbifold effective theory) based on one's choice of $C_{IJK}$ components (as this is responsible for determining the K\"{a}hler scalar manifold found from orbifolding the YMESGT sector).  

We can abandon the restriction of minimal field content (i.e., relax the choices made in appendix C.1) if non-trivial $C_{IJK}$ are desired.     

That being said, our first focus in this paper is on obtaining a hypermultiplet field content used in orbifold-GUTs, so we will assume one of the choices in appendix C.1 has been made for the hypermultiplet discussion.  Of course, once we move on to non-compact gaugings and tensor couplings, we will not be assuming a particular form for $C_{IJK}$.  Now that we know where our gauge fields are coming from, it may be desirable to have $5D$ fermions and scalars in non-singlet representations of the gauge group as in orbifold-GUTs.  There are two avenues here:
\begin{itemize}
\item
Couple hypermultiplets to the simple MESGT described in appendix C.1 (or some other MESGT if desired); this is the discussion of section 3.2, or
\item
Consider a new MESGT with non-singlets in the decomposition under the desired gauge group $K$ (in which case we must deal with tensor fields and at least one abelian factor in the gauge group); such is the discussion of section 4.  
\end{itemize}

\subsection{YMESGTs coupled to $n_{H}$ hypermultiplets}
    
The coupling of $n_{H}$ hypermultiplets in $5D$ $\mathcal{N}=2$ supergravity is briefly discussed in appendix B (the main reference being~\cite{CDA:00apr}).  The scalar fields in the $n_{H}$ hypermultiplets parametrize a quaternionic manifold, $\mathcal{M}_{Q}$, of real dimension $4n_{H}$ (the quaternionic dimension is $\mbox{dim}_{\mathbb{H}}(\mathcal{M}_{Q})=n_{H})$~\cite{BW:83}.  The total scalar manifold of a MESGT coupled to hypermultiplets is then  
\[
\mathcal{M}\equiv \mathcal{M}_{R}\times \mathcal{M}_{Q},
\]       
with isometry group 
$Iso(\mathcal{M})\simeq Iso(\mathcal{M}_{R}) \times Iso(\mathcal{M}_{Q})$~\cite{S:85}.  
 
Once again, one can gauge a subgroup $K\subset G\subset Iso(\mathcal{M})$.  In particular, since we want non-trivially charged hypermultiplets, $K\subset Iso(\mathcal{M}_{R})\times Iso(\mathcal{M}_{Q})$, where $K$ is non-abelian and simple, or including an abelian factor.  For the non-abelian part, $K\subset G_{1}\times G_{2} \subset Iso(\mathcal{M})$ such that $K$ is isomorphic to both a subgroup of $G_{1}\subset Iso(\mathcal{M}_{R})$ and a subgroup of $G_{2}\subset Iso(\mathcal{M}_{Q})$.  Again, if the abelian factor consists of an isometry of $\mathcal{M}_{R}$, we must have tensors charged under at least this factor.  If it consists only of an isometry of $\mathcal{M}_{Q}$, no charged tensors are present. 
      
Due to supersymmetry, the scalars in $5D$ hypermultiplets must generally form the 
\[
2(\Sigma_{i}\mathbf{n_{i}})\oplus 2\Sigma_{\alpha}(\mathbf{n_{\alpha}}\oplus \mathbf{\bar{n}_{\alpha}})
\]
representation of any gauge group we consider, where $i$ labels pseudoreal irreps and $\alpha$ labels real and complex irreps.  Notice the factor of two (not four) in front of the pseudoreal terms; this is due to the fact that such $5D$ hypermultiplets can be split into two $4D$ chiral multiplets each in a pseudoreal irrep of a gauge group, which is a reflection of the fact that they are self-conjugate.  The main groups of concern here ($SU(5)$, $SO(10)$, and $E_{6}$) do not carry pseudoreal representations.  However, $E_{7}$ has the $\mathbf{56}$, which will play a role later on.

Let us set our notation.  A $5D$ $\mathcal{N}=2$ hypermultiplet contains four real scalars.  If $4m$ scalars form the real or complex representations $\mathbf{m}_{\mathbb{H}}\equiv 2(\mathbf{m}\oplus\mathbf{\bar{m}})$, we say that the hypermultiplet is in the $\mathbf{m}$ (we do not need to distinguish from $\mathbf{\bar{m}}$).  If the $4m$ scalars form the pseudoreal $\mathbf{m}_{H}=2[\mathbf{2m}]$, we will denote the hypermultiplet as being in the $\mathbf{2m}$.  A $4D$ $\mathcal{N}=1$ spin-1/2 multiplet contains two real scalars.  If $2m$ scalars form the real or complex $\mathbf{m}_{\mathbb{C}}\equiv \mathbf{m}\oplus\mathbf{\overline{m}}$, we say that the spin-1/2 multiplet is in the $\mathbf{m}$ (or equivalently, there is a chiral plet in the $\mathbf{m}$ and its CPT conjugate); while if the $2m$ scalars form the $\mathbf{2m}$, we say the spin-1/2 multiplet is in the pseudoreal $\mathbf{2m}$.    

To list the hypermultiplet content allowed by supergravity, one starts with a particular quaternionic scalar manifold admitting the desired gauge group and corresponding to some number of hypermultiplets.  The scalars transform in the above representation of the maximal compact subgroup of $Iso(\mathcal{M}_{Q})$.  The representations break down under the global symmetry group of the Lagrangian, $G\subset Iso(\mathcal{M}_{R})\times Iso(\mathcal{M}_{Q})$.  Finally, under the group we wish to gauge, $K\subset G$, the scalars decompose further; the spectrum of hypermultiplet representations in the theory.                  

It is a simple exercise to write down the list of possible matter representations given a gauge group and quaternionic scalar manifold.  We first list the hypermultiplets that appear in theories based on homogeneous/symmetric spaces, followed by a brief discussion of those based on homogeneous/non-symmetric quaternionic scalar manifolds.  We then make comments on non-homogeneous quaternionic scalar manifolds in section 3.5.  
\newpage
\subsection{Theories based on homogeneous quaternionic scalar manifolds}

Homogeneous spaces are characterized by the fact that the isometry group acts transitively on the space $\mathcal{M}$.  Such spaces are isomorphic to $Iso(\mathcal{M})/H$, where $H$ is the isotropy group of $\mathcal{M}$.  Generally, though not always, homogeneous spaces are the ones admiting large isometry groups, which can then admit large gauge groups.  A listing of the homogeneous ``special" quaternionic manifolds appearing in the coupling to supergravity (in $3,4,$ and $5$ spacetime dimensions) can be found in~\cite{dWVP:95may}; the spaces we are interested in are given in table 1.  For these theories, $H$ is the symmetry group $G$ of the Lagrangian.    
\begin{center}
\begin{large}
\begin{tabular}{|c|c|c|c|}  \hline
Type & Scalar Manifold & $\mbox{dim}_{\mathbb{H}}(\mathcal{M}_{Q})$ & $H_{Q}$-rep of scalars \\ \hline \hline 
 & & & \\
$L(0,P)$ & $\frac{SO(P+4,4)}{SO(P+4)\times SO(4)}$ & $P+4$ & $(\mathbf{P+4},\mathbf{4})$\\ 
 & & & \\ \hline
 & & & \\
$L(2,1)$ & $\frac{E_{6}}{SU(6)\times SU(2)}$ & $10$ & $(\mathbf{20},\mathbf{2})$ \\ 
 & & & \\ \hline
 & & & \\
$L(4,1)$ & $\frac{E_{7}}{\overline{SO(12)}\times SU(2)}$ & $16$ & $(\mathbf{32'},\mathbf{2})$\\ 
 & & & \\ \hline
 & & & \\
$L(8,1)$ & $\frac{E_{8}}{E_{7}\times SU(2)}$ & $28$ & $(\mathbf{56},\mathbf{2})$\\ 
 & & & \\ \hline
 & & & \\
$L(-3,P)$ & $\frac{USp(2P+2,2)}{USp(2P+2)\times SU(2)}$ & $P+1$ & $(\mathbf{2P+2},\mathbf{2})$\\ 
 & & &  \\ \hline
 & & & \\
$L(-2,P)$ & $\frac{SU(P+2,2)}{SU(P+2)\times SU(2)\times U(1)}$ & $P+2$ & $(\mathbf{P+2},\mathbf{2})\oplus (\mathbf{\overline{P+2}},\mathbf{2})$ \\ 
 & & & \\ \hline
 & \multicolumn{3}{c|}{} \\
$L(q,P)$ & \multicolumn{3}{c|}{Discussed in text} \\ \hline 
\end{tabular} \vspace{4mm}\\ \end{large}
Table 1: List of relevant homogeneous \textit{quaternionic} scalar manifolds for the discussion in the paper.  The ``type" of space is the classification name as in~\cite{dWVP:95may}; the number of real scalar fields is $4\, \mbox{dim}_{\mathbb{H}}(\mathcal{M}_{Q})$; the scalars form reps of the isotropy group $H_{Q}$ as listed.
\end{center}
\newpage

\newpage
After reducing the scalars of these theories down to reps of $E_{6}$, $SO(10)$, and $SU(5)$, we find the possible hypermultiplet representations charged under those gauge groups; they are given in tables 2, 3 and 4, respectively.  We have listed those cases with the lowest irrep dimensions (with the exception of the case $L(0,74)$).
\vspace{4mm}\\
\begin{center}
\begin{tabular}{|ccccc|} \hline
Type & & $\mbox{dim}_{\mathbb{H}}(\mathcal{M}_{Q})$ & & $K$-rep of hypermultiplets \\ \hline \hline
$L(-2,P)$ & & $P+2=27n$ & & $n(\mathbf{27})$\\ \hline
$L(0,P)$ & & $P+4=54$ & & $\mathbf{27}\oplus\mathbf{27}$\\
 & & $78$ & & $\mathbf{78}$\\ \hline
$L(8,1)$  & & $28$ & & $\mathbf{1}\oplus \mathbf{27}$ \\ \hline
\end{tabular}
\vspace{2mm} \\
Table 2: List of hypermultiplets in lowest dimensional representations when gauging $E_{6}$.\\  
In the table, $n=1,2,\ldots$.
\vspace{10mm}\\
\begin{tabular}{|ccccc|} \hline
Type & & $\mbox{dim}_{\mathbb{H}}(\mathcal{M}_{Q})$ & & $K$-rep of hypermultiplets \\ \hline \hline
$L(-2,P)$ & & $P+2=27n$ & & $n(\mathbf{1}\oplus \mathbf{10}\oplus\mathbf{16})$ \\ \hline
$L(0,P)$ & & $P+4=10n$ & & $n(\mathbf{10})$ \\
 & & $78$ & & $\mathbf{1}\oplus 2(\mathbf{16})\oplus \mathbf{45}$ \\ \hline
$L(4,1)$ & & $16$ & & $\mathbf{16}$ \\ \hline
$L(8,1)$ & & $28$ & & $2(\mathbf{1})\oplus \mathbf{10}\oplus \mathbf{16}$ \\ \hline
\end{tabular}
\vspace{2mm} \\
Table 3: List of hypermultiplets in lowest dimensional representations when gauging $SO(10)$.\\
  In the table, $n=1,2,\ldots$.
\vspace{10mm} \\
\begin{tabular}{|ccc|} \hline
Type &  $\mbox{dim}_{\mathbb{H}}(\mathcal{M}_{Q})$ &  $K$-rep of hypermultiplets \\ \hline \hline
$L(-3,P)$ &  $P+1=5n$ &  $n(\mathbf{5})$\\ \hline
$L(-2,P)$ &  $P+2=5n$ &  $n(\mathbf{5})$ \\ 
 &  $27n$ &  $2n(\mathbf{1})\oplus 3n(\mathbf{5})\oplus n(\mathbf{10})$ \\ \hline
$L(0,P)$ & $P+4=10n$ & $2n(\mathbf{5})$\\
  & $78$ & $4(\mathbf{1})\oplus 2(\mathbf{5})\oplus 4(\mathbf{10})\oplus \mathbf{24}$ \\ \hline
$L(2,1)$ & $10$ & $\mathbf{10}$ \\ \hline
$L(4,1)$ & $16$ & $\mathbf{1}\oplus \mathbf{5}\oplus \mathbf{10}$ \\ \hline
$L(8,1)$ & $28$ & $3(\mathbf{1})\oplus 3(\mathbf{5})\oplus \mathbf{10}$   \\ \hline
\end{tabular}
\vspace{2mm} \\
Table 4: List of hypermultiplets in lowest dimensional representations when gauging $SU(5)$. \\
In the table, $n=1,2,\ldots$.
\end{center}
\newpage

\subsection{Homogeneous, non-symmetric spaces}

$\underline{\textbf{Other}\;\mathbf{L(q,P)}}$: 

We will now go through the type of hypermultiplet content obtained by coupling homogeneous/non-symmetric quaternionic scalar mnaifolds to supergravity.  There will generically be a number of gauge singlets in addition to non-trivial irreps.  
The isotropy group for quaternionic scalar manifolds $L(q,P)$ that are homogeneous, but non-symmetric is
\[
H=SO(q+3)\times SU(2)\times S_{q}(P,\dot{P}),
\]
where $S_{q}(P,\dot{P})$ is given in table 10 of~\cite{dWVP:95may}.  The quaternionic dimension of the manifold is 
$$n+1=4+q+(P+\dot{P})D_{q+1}.$$ 
The isometry algebra has a three-grading with respect to a generator $\epsilon'$:
\begin{gather}
V=V_{0}\oplus V_{1}\oplus V_{2} \nonumber\\
V_{0}=\epsilon' \oplus so(q+3,3) \oplus s_{q}(P,\dot{P}) \nonumber\\
V_{1}=(1,spinor,vector) \nonumber\\
V_{2}=(2,vector,0), \nonumber
\end{gather}
where $spinor$ is the spinor representation of $so(q+3,3)$, which is dimension $4D_{q+1}$. 
\begin{itemize}
\item
$\mathbf{SU(5)\subset H}$\\
The only way this can arise is when $S_{q}\equiv U(5)$; in turn, this occurs for:
\begin{center}
\begin{tabular}{cccc}
$q$ & $D_{q+1}$ & $P$ & $dim_{\mathbb{H}}(\mathcal{M}_{Q})$\\ \hline
$2$ & $4$ & $5$ & $26$\\ 
$6$ & $16$ & $5$ & $90$
\end{tabular}
\end{center}
When we gauge $SU(5)$, the scalars will then have the following representation under the gauge group:
\[
[ \mathbf{1}\oplus (9+3q)\mathbf{1}] \oplus [2D_{q+1}(\mathbf{5}\oplus\mathbf{\bar{5}})] \oplus [(q+6)\mathbf{1}].
\]

\item
$\mathbf{SU(6)\subset H}$\\
Again, the only way to get this is to have $S_{q}\equiv U(6).$  This case is then similar to the above, with $P=6$.  We will get vectors and singlets of $SU(6)$, and therefore $\mathbf{5}$s and singlets of $SU(5)$. 
\item
$\mathbf{SO(10)\subset H}$\\
\textbf{A.}  $\mathbf{S_{q}\equiv SO(10)}$\\
The choices are then
\begin{center}
\begin{tabular}{cccc}
$q$ & $D_{q+1}$ & $P$ & $dim_{\mathbb{H}}(\mathcal{M}_{Q})$\\ \hline
$-1$ & $1$ & $10$ & $13$\\
$1$ & $2$ & $10$ & $25$\\
$7$ & $16$ & $10$ & $171$\\
\end{tabular}  
\end{center}
The scalars in these cases form the following representation under $SO(10)$:
\[
[\mathbf{1}\oplus (3q+9)\mathbf{1})]\oplus [(4D_{q+1})\mathbf{10}]\oplus [(q+6)\mathbf{1}]
\]
\textbf{B.}  $\mathbf{SO(q+3)\equiv SO(10)}$ ($q=7$, $D_{q+1}=16$, $P=$arbitrary)\\
These spaces have quaternionc dimension $11+16P$.  However, under an $SO(10)$ subgroup of the isotropy group, the scalars form a set of representations inconsistent with supersymmetry, as they do not form quaternions that can sit in $5D$ hypermultiplets.  Thus, neither $SO(10)$ nor its $SU(5)$ subgroup can be consistently gauged.  
\item $\mathbf{Sp(10)\subset H}$  
This case is similar to the above and will not be discussed.  
      
\end{itemize}

\subsection{Comments on theories based on non-homogeneous spaces}

Non-homogeneous real and quaternionic scalar manifolds are relevant in string compactifications.  For example, it has been shown that, in the special case of the universal hypermultiplet of string compactifications, the quaternionic scalar manifold generally becomes non-homogeneous after string corrections are considered~\cite{AS:97jun}.  In string theory on a Calabi-Yau manifold, 1-loop effects can show up in $11D$ supergravity on a Calabi-Yau, and thus can appear in compactifications to $5D$.  However, non-homogeneous quaternionic manifolds have not been generally classified; in particular those admiting relatively large isometry groups (suitable for obtaining large gauge groups with charged hypermultiplets in phenomenologically interesting representations).  We can list three methods from the literature for obtaining (non-compact) non-homogeneous quaternionic manifolds.\\  
\textbf{(i)}
In $5D$, it has been shown that there are Maxwell-Einstein supergravity theories with large isometry groups based on real non-homogeneous manifolds parametrized by scalars from the vector multiplets~\cite{EGZ:01aug, GZ:03apr}.  By dimensionally reducing these five-dimensional theories to three dimensions, we can obtain non-homogeneous quaternionic scalar manifolds with large isometry groups, which we can couple to $5D$ supergravity.  Such a reduction was done for the theories with a special class of symmetric scalar manifolds~\cite{GST:83}, and later an analysis for more general homogeneous spaces appeared~\cite{dWVP:95may}.  An analysis of the isometries of the non-homogeneous quaternionic spaces arising from the theories in~\cite{GZ:03apr} is a work in progress by the authors of~\cite{GMZ:05}.\\     
\textbf{(ii)}
One may construct $4n+4$-dimensional non-homogeneous quaternionic manifolds $\hat{\mathcal{M}}$ by fibering over an arbitrary $4n$-dimensional quaternionic base manifold $\mathcal{M}$ with isometry group $Iso(\mathcal{M})$, as discussed in~\cite{PP:86}.  The isometry group is locally $Iso(\mathcal{M})\times SU(2)$.
\textbf{(iii)}
In ~\cite{KG:87}, it was shown how $4n$-dimensional generalizations of the four-dimensional non-homogeneous quaternionic space of Pedersen~\cite{HP:87} (originally considered by Hitchin) could be constructed.  These spaces have $SU(n)\times SU(2)\times U(1)$ isometries, and seem to be non-homogeneous forms of the space $L(-2,P)$ in table 1.  Aside from the spaces that are cosets of exceptional groups, most of the infinite families of quaternionic manifolds classified to date are quaternionic quotients by quaternionic isometries of the quaternionic projective space $\mathbb{H}H^{n}=Sp(2n+2)/Sp(2n)\times SU(2)$, or non-compact or pseudo-quaternionic forms thereof.  The spaces in~\cite{KG:87} are of this type, and are presumably pseudo-quaternionic analogues of the spaces $L(-2,P)$ in table 1.      
     
We will not attempt to discuss the possible roles of these theories within this paper.

\subsection{Summary and discussion: YMESGTs coupled to hypermultiplets}

Just as all compact gaugings are possible in pure MESGTs, all compact gaugings are possible when coupling to non-trivially charged hypermultiplets.  For example, the theories of type $L(0,P)$ admit any compact gauge group $K$, since the adjoint representation of $K$ can always be embedded in the fundamental representation of $SO(P+4)$ if $P+4\geq \mbox{dim}(K)$.  However, the resulting hypermultiplet content may be undesirable.  This then restricts the number of theories in which we can obtain both the gauge group and hypermultiplet content desired (a theory being uniquely determined by the scalar manifold up to possible arbitrary parameters).   

The restriction is even more severe than this.  It is not guaranteed that an arbitrary set of hypermultiplets can be obtained by finding a suitable \textit{quaternionic} manifold admiting the desired gauging.  This is clear at least within the set of quaternionic scalar manifolds that are homogeneous, as discussed in this paper.  For example, one may not obtain an arbitrary number of hypermultiplets in the $\mathbf{10}$ of an $SU(5)$ gauge group.  It should be noted that one may \textit{not} get around this restriction by simply coupling two quaternionic scalar manifolds $\mathcal{M}_{1}$ and $\mathcal{M}_{2}$ such that $\mathcal{M}_{Q}=\mathcal{M}_{1}\times \mathcal{M}_{2}$, since these are no longer \textit{quaternionic} manifolds.  A quaternionic structure is necessary for coupling to supergravity~\cite{BW:83}.  

However, there is a way to construct new quaternionic manifolds from a pair $(\mathcal{M}_{1},\mathcal{M}_{2})$ of quaternionic manifolds, as discussed in~\cite{AS:91}.  The construction relies on the fact that a $4n$-dimensional hyper-K\"{a}hler manifold ($Sp(2n)$ holonomy) can be constructed as a bundle over a $(4n-4)$-dimensional quaternionic manifold ($Sp(2n-2)\times SU(2)$ holonomy).  Let $\mathcal{M}_{1}$, $\mathcal{M}_{2}$ be quaternionic scalar manifolds of dimension $4n_{1}$ and $4n_{2}$, respectively.  Then there exists a $4n_{1}+4n_{2}+4$-dimensional quaternionic manifold $\mathcal{J}(\mathcal{M}_{1},\mathcal{M}_{2})$ called the ``quaternionic join".  Let $U_{1},\,U_{2}$ be the hyper-K\"{a}hler bundles with base manifolds $\mathcal{M}_{1}$ and $\mathcal{M}_{2}$, respectively.  The hyper-K\"{a}hler manifold $U_{1}\times U_{2}$ is then the bundle over $\mathcal{J}(\mathcal{M}_{1},\mathcal{M}_{2})$.  Locally, the manifold $\mathcal{J}(\mathcal{M}_{1},\mathcal{M}_{2})$ is a $\mathbb{Z}_{2}$ quotient of $\mathcal{M}_{1}\times \mathcal{M}_{2}$.  The construction requires hyper-K\"{a}hler manifolds admitting a hyper-K\"{a}hler potential (a K\"{a}hler potential for each of the three complex structures).  The global manifold does not need to carry the isometries of $\mathcal{M}_{1}$ or $\mathcal{M}_{2}$, which may ruin the options for gauging.  Even if a particular gauge group is still allowed, the local structure does not necessarily admit the representation $R[\mathcal{M}_{1}]\oplus R[\mathcal{M}_{2}]\oplus \mathbf{1}$, where $R[\mathcal{M}]$ is the representation of the scalars (parametrizing $\mathcal{M}$) under the gauge group.  This is in contrast with the case of vector and tensor scalars, which locally form a product structure $\mathcal{M}_{V}\times\mathcal{M}_{T}$ so that in any neighborhood, scalars can always be divided up into $R[\mathcal{M}_{V}]\oplus R[\mathcal{M}_{T}]$; i.e., representations of a ``vector sector" and reps of a ``tensor sector". \vspace{2mm}\\
\textbf{Theories with only bulk Higgs coupling} 

In many phenomenological models, the bulk theory is a super-Yang-Mills theory coupled to bulk Higgs hypermultiplets in particular representations of the gauge group.  It is simple to obtain such field content in supergravity.  In the case of $SU(5)$ gauging, one can couple any number $n$ of hypermultiplets in the $\mathbf{5}$ by coupling the scalar manifolds $L(-3,5n-1)$ or $L(-2,5n+2)$.  In the case of $SO(10)$ gauging, one can couple any number $n$ of hypermultiplets in the $\mathbf{10}$ by coupling the scalar manifold $L(0,10n-4)$.  In gauging $E_{6}$, any number $n$ of hypermultiplets in the $\mathbf{27}$ may be obtained by coupling to the scalar manifold $L(-2,27n-2)$.\vspace{2mm}
\textbf{Theories with bulk matter}

As an interesting illustration of the options that supergravity allows (within the homogeneous quaternionic cases), it appears that to obtain a generation of bulk matter hypermultiplets in the $\mathbf{5}\oplus\mathbf{10}$ of $SU(5)$ one \textit{must} include a gauge singlet hypermultiplet (see table 4).  But the corresponding theory of type $L(4,1)$ naturally allows gauging of $SO(10)$, under which the hypermultiplets form the irreducible $\mathbf{16}$ (see table 3).  One can then orbifold one of these theories as desired.  

If a Higgs sector and single generation of matter hypermultiplets is to be coupled in the bulk $SU(5)$ theory, one \textit{must} add two or three additional singlets, which corresponds to the coupling of a different scalar manifold: $L(-2,25)$ or $L(8,1)$, respectively (see table 3).  Now the field content can sit in smaller set of reps if we gauge $E_{6}$ instead, under which the hypermultiplets form a single $\mathbf{27}$ or $\mathbf{27}\oplus\mathbf{1}$, respectively (see table 2).  Again, one may then orbifold one of these theories as desired.  One can go further.  The hypermultiplets in the $\mathbf{27}\oplus\mathbf{1}$ in the $L(8,1)$ theory can form the $\mathbf{56}$ pseudoreal irrep if we gauge the $E_{7}$ allowed by that space.  Upon orbifolding the $E_{6}$ theory, we would get a chiral multiplet in the $\mathbf{27}\oplus\mathbf{1}$ and its CP conjugate.  Orbifolding the $E_{7}$ theory yields a spin-1/2 multiplet in the self-conjugate $\mathbf{56}$, which is not good phenomenologically.

Finally, suppose one desires three generations of matter and Higgs in an $SU(5)$ theory.  Once again, one must couple two additional singlet hypermultiplets for each generation, corresponding to the scalar manifold $L(-2,79)$.  This begs the question why we shouldn't gauge $E_{6}$ instead such that the fields form three generations of $\mathbf{27}$.

One might instead envision the breaking \textit{in five dimensions} of $SO(10)$ to $SU(5)$ gauge group in the theory with scalar manifold $L(4,1)$.  If spontaneous, this breaking would require an additional $5D$ Higgs sector, which would require a coupling of a \textit{different} quaternionic scalar manifold.  For example, one would have to couple to $L(-2,25)$ or $L(8,1)$, which introduce additional gauge singlets.  Note that if the group is broken to $SU(5)\times U(1)$, the $U(1)$ factor consists of mixture of gauged real and quaternionic manifold isometries (since all of the $SO(10)$ gauge symmetries are gauged isometries of \textit{both} the real and quaternionic scalar manifolds).  However, tensors charged under this $U(1)$ are not required since there are massive vector multiplets from the Higgs mechanism that take their place.   
  
Similarly, we could evisage the spontaneous breaking of the above $E_{6}$ theory to $SO(10)$.  We'd need to couple a new scalar manifold admitting a Higgs sector (the simplest being $L(-2,52)$ or $L(0,50)$ with two $\mathbf{27}s$), but this would make the field content more complicated in the end.  

Alternatively, the $5D$ breaking could be performed by Wilson lines of, e.g., the $U(1)$ factor in the subgroup $SU(5)\times U(1)\subset SO(10)$ or of $SO(10)\times U(1)\subset E_{6}$.  Some details of such breaking and the relation to parity assignments is discussed in~\cite{SM3}; for related issues of boundary conditions and Wilson lines in $S{1}/\Gamma$ orbifold field theory scenarios, see~\cite{wilson}.\vspace{2mm}\\
\textbf{Partial unification in four dimensions}                           

Most of our discussion is in the context of orbifold-GUTs wherein the Standard Model gauge group remains at one of the fixed points.  If one wants to break to a partially unified group at one of the fixed points, the Higgs content in the bulk must of course be enlarged (suppose that we do not wish to turn to string theory and its twisted sector states). 
\vspace{2mm}\\
$\mathbf{SO(10)\rightarrow SU(4)\times SU(2)_{L}\times SU(2)_{R}}$

In four dimensions, we would need the $\mathbf{16}\oplus\mathbf{\overline{16}}$ of $SO(10)$ to break the Pati-Salam gauge group (PS) to that of the SM (we need the $(4,1,2)$ of PS and its conjugate to perform the breaking), as well as the electroweak breaking.  Since each hypermultiplet can provide a $4D$ left-chiral $\mathbf{16}$ \textit{or} $\mathbf{\overline{16}}$, we will need to have 2 five-dimensional hypermultiplets in the $\mathbf{16}$.  In addition, a single generation of matter is in the $\mathbf{16}$ of $SO(10)$.  The minimal way to get multiple $\mathbf{16}$s in the bulk is by coupling an $SO(10)$ YMESGT to hypermultiplets whose scalars parametrize the manifold $L(-2,27n-2)$; that is, we must have $n$ hypers in the $\mathbf{1}\oplus \mathbf{10}\oplus \mathbf{16}$.  Therefore, extra fields must come along ($n$ copies of $(2,2,1)\oplus (1,1,6)\oplus (1,1,1)$ of $SU(2)\times SU(2)\times SU(4)$).  These extra states will show up as color triplets and extra weak doublets and singlets, and must be made massive via boundary conditions.  This in turn will affect the gauge coupling running, possibly adversely.  Anyway, the necessary additions beg the question: why not gauge $E_{6}$ instead so that the fields form an irrep?\vspace{2mm}\\
$\mathbf{E_{6}\rightarrow SU(3)_{c}\times SU(3)_{L}\times SU(3)_{R}}$

To round out the discussion, suppose we wish to have an $SU(3)^{3}$ trinification~\cite{trinification} scenario at the fixed points of the orbifold.  In four dimensions, we would need the $\mathbf{27}\oplus\mathbf{\overline{27}}$ of $E_{6}$ to break the trinification group (TG) to SM, and then down to the visible $SU(3)_{c}\times U(1)_{em}$.  Since the color singlets inside the $\mathbf{27}$s are used to perform gauge symmetry breaking, we must add an additional $\mathbf{27}$ for each generation of matter.  We can obtain such a model by coupling an $E_{6}$ YMESGT to $n$ hypermultiplets whose scalars once again parametrize the manifold $L(-2,27n-2)$.

\section{YMESGTs coupled to tensor multiplets}

As mentioned previously, we may gauge any compact group in the framework of $5D$ $\mathcal{N}=2$ supergravity, but we cannot presume any charged field content we like.  A real Riemannian scalar manifold must be specified for which $K\subset G\subset Iso(\mathcal{M}_{R})$.  In~\cite{TPvN:84}, it was shown that tensor multiplets carrying non-trivial representations of a group $G$ and satisfying a self-duality condition are required to come in complex conjugate pairs $\Sigma_{i}\mathbf{m}_{i}\oplus \Sigma_{i}\mathbf{\bar{m}}_{i}$.  However, it was shown in~\cite{GZ:99dec} that, when gauging $K$, the complex $K$-representations should be of ``quaternionic type"\footnote{``Quaternionic type" meaning that there exists a $K$-invariant anti-symmetric bilinear form on the vector space.}; that is, symplectic representations.  Tensor multiplets charged under a gauge group $K$ therefore arise when there are symplectic, non-singlet representations of $K$ in the decomposition appearing in eqn(3.1).  It is sufficient to find complex representations of a compact gauge group $K_{semi-simple}\times U(1)$ coming in pairs $\mathbf{m}\oplus \mathbf{\bar{m}}$. 

We may now look for gauge theories we are interested in that admit tensor couplings.  We will ignore any coupling to hypermultiplets here so that the scalar manifold is $\mathcal{M}_{R}$, and we can gauge $K\subset G\subset Iso(\mathcal{M}_{R})$, where $G$ consists of isometries that are symmetries of the Lagrangian.  In appendix A, it is recalled that a rank 3 symmetric tensor $C_{IJK}$ uniquely determines the form of a MESGT (up to possible arbitrary reparametrizations).  Thus, the non-trivial invariance group of this tensor is precisely $G$.  An algebraic analysis of the form of this tensor is useful for understanding the vector and tensor field content in a YMESGT, and for gauging purposes is equivalent to the geometric analysis of the form of the scalar manifold and its isometry group.  Note that we are now abandoning the simple choice made in appendix C.1 wherein $C_{ijk}=0$.      

It was shown in~\cite{GZ:99dec} that for theories based on \textit{homogeneous, symmetric} real scalar manifolds, there aren't any large non-abelian gauge groups (such as the typical GUT groups) with charged tensors.    

\subsection{Homogeneous, non-symmetric scalar manifolds}
These scalar manifolds are described in~\cite{dWVP:95may}.  The presence of tensor fields in these theories was addressed in~\cite{GZ:99dec}, though we perform a slightly different analysis.  

The isometry algebra has a one-grading with respect to a generator $\lambda$:
\[
\begin{split}
\chi=\chi_{0}\oplus \chi_{3/2}\\
\chi_{0}=\lambda \oplus so(q+1,1) \oplus s_{q}(P,\dot{P})\\
\chi_{3/2}=(spinor,vector),
\end{split}
\]
where $spinor$ denotes the spinor representation of $SO(q+1,1)$ of dimension $D_{q+1}$, and the groups corresponding to the algebras $s_{q}(P)$ are listed in~\cite{dWVP:95may}.  The isotropy group is 
\[
H=SO(q+1)\times S_{q}(P,\dot{P}),
\]  
where $S_{q}(P,\dot{P})$ is the group corresponding to the algebra $s_{q}(P,\dot{P})$.  The dimension of the real Riemannian scalar manifold is $(n-1)=2+q+D_{q+1}P.$  The vector multiplets form the following representation under $H$:
\begin{equation}
\mathbf{(1,1)}\oplus \mathbf{(q+1,1)}\oplus \mathbf{(D_{q+1},P)}.
\end{equation}

The condition for gauging a group $K$ is that the adjoint representation of $K$ should appear in the decomposition into $K$-reps (see eqn(3.1)). 
  
\begin{itemize}
\item
$K\subset SO(q+1)$ and $q\geq 9$, $D_{q+1}\geq 32$, $P=$arbitrary

$\mathbf{K=SU(5)}$\\
\textbf{(i)}  If $10\leq q+1 \leq 23$, then the only possibility for the adjoint rep of $K$ to exist is if the spinor rep of $SO(q+1,1)$ contains it in the decomposition.  The spinor rep is of dimension $32\leq D_{q+1}\leq 4096$ in our special case.  It seems that the largest representation dimension decomposing from the spinor representation of $SO(q+1,1)$ is the $\mathbf{16}$ of $SO(10)$ so that we cannot gauge $SU(5)$.\\
\textbf{(ii)}  If $q+1\geq 24$, any value of $q$ in this range allows gauging of $SU(5)$, with $(q+1)-24$ spectator fields.  The remaining vector fields are $P$ copies of the decomposition of $spinor[SO(q+1,1)]$ into $SU(5)$-reps.  This yields a large number of tensor fields ($>4096$).  

$\mathbf{K=SO(10)}$\\
\textbf{(i)}  If $10\leq q+1\leq 44$, the only possibility for the adjoint rep of $K$ to exist is if the spinor rep of $SO(q+1,1)$ contains it in the decomposition.  The spinor rep is of dimension $32\leq D_{q+1}\leq 2^{23}$, but the largest irrep in the decomposition into $SO(10)$ representations is the $\mathbf{16}$.  Therefore, $SO(10)$ cannot be gauged for these values of $q$.\\
\textbf{(ii)}  If $q+1\geq 45$, we can gauge $SO(10)$ for all values of $q$ in this range.  The $(q+1)-45$ vector fields not involved in this gauging are spectators.  The remaining vector fields are in $P$ copies of the decomposition of $spinor[SO(q+1,1)]$ into $SO(10)$-reps.  This yields a large number of tensor fields ($>2^{23}$) in the $\mathbf{16}\oplus\mathbf{\overline{16}}$.
    
\item
$K\subset S_{q}(P)\equiv SO(P)$ with $P\geq 10$\\
We require that $adj(K)\subset vector[S_{q}(P)]$.  The adjoint of any compact Lie group sits in the $\mathbf{n}$ of $SO(n)$ if $n \geq \mbox{dim}[adj(K)]$.  There will then be $n-\mbox{dim}[adj(K)]$ spectator vector fields in addition to the graviphoton.  Therefore, no tensors charged under a non-abelian gauge group appear. 

\item
$K\subset S_{q}(P)\equiv U(P)$ with $P\geq 5$\\
The values of $q$ are restricted to $q=2\;\mbox{mod}(8)$ and $6\;\mbox{mod}(8)$. 
Again, we require $adj(K)\subset vector[S_{q}(P)]$.  The vector fields form the $\mathbf{1}\oplus (q+1)\mathbf{1}\oplus (D_{q+1})\mathbf{P}$ of $S_{q}(P)$.  For the adjoint of $U(5)$ to appear, we require $P\geq 25$.  However, there aren't any tensors to be charged with respect to the $U(1)$ factor.  This applies for other non-abelian cases.

\end{itemize}

\subsection{Non-homogeneous scalar manifolds}

\begin{itemize}
\item
Theory with $C_{ijk}=0$\\
This is the choice in case \textbf{1} of appendix C.1.  As is clear from the discussion in appendix C.2, these theories do not admit tensor multiplets charged under non-abelian gauge groups.\\
\end{itemize}
A set of interesting theories are based on Lorentzian Jordan algebras~\cite{GZ:03apr}\footnote{These were originally called ``Minkowski" Jordan algebras in that work.} $J^{\mathbb{A}}_{(1,N)}$ of degree $(N+1)$, where $\mathbb{A}=\mathbb{R},\mathbb{C},\mathbb{H}$; there is also the exceptional theory based on $J^{\mathbb{O}}_{(1,2)}$.  These theories are listed below, where $G$ denotes the invariance group of the Lagrangian.  In constrast to the theories with homogeneous scalar manifolds, these theories admit GUT groups coupled to non-trivially charged tensor multiplets.  For phenomenological reasons, though, we focus on $SU(5)\times U(1)$ gauging.  
\vspace{2mm} 
\begin{center}
\begin{tabular}{|c|c|c|} \hline
J.A. & $\mbox{dim}(\mathcal{M}_{R})$ & $G$  \\ \hline
$J^{\mathbb{R}}_{(1,N)}$ & $N(N+3)/2-1$ & $SO(N,1)$ \\
$J^{\mathbb{C}}_{(1,N)}$ & $N(N+2)-1$ & $SU(N,1)$ \\
$J^{\mathbb{H}}_{(1,N)}$ & $N(2N+3)-1$ & $USp(2N,2)$ \\
$J^{\mathbb{O}}_{(1,2)}$ & $25$ & $F_{(4,-20)}$ \\
\hline
\end{tabular}
\end{center}
\vspace{2mm}
These theories are all examples of ``\textbf{unified}" MESGTs, in the sense that there is a continuous symmetry connecting every field of the theory:\footnote{We will use boldface to distinguish this from any other sense of ``unified".}
\[
\begin{array}{ccccc}
\leftrightarrow & \leftrightarrow & \leftrightarrow & & \\
\{e^{m}_{\mu} & \Psi_{\mu} & A^{0}_{\mu}\} & & \\
 & & \Updownarrow & & \\
 & & \{A^{i}_{\mu} & \lambda^{a} & \phi^{x}\}\\
 & & \leftrightarrow & \leftrightarrow & \leftrightarrow
\end{array}
\]
where horizontal arrows represent (local) supersymmetry action, and the vertical arrow represents the action of a \textit{simple} global symmetry group $G$, like those listed in the above theories.  (For more on \textbf{unified} MESGTs, see~\cite{GST:84, GZ:03apr}.)

A general discussion of the tensor couplings in these theories can be found in~\cite{GZ:03apr}, which we use to write down the theories of interest to us here.
\begin{itemize}
\item
$J^{\mathbb{R}}_{(1,N)}$\\
One can gauge $SU(n)\times U(1)\subset SO(2n,1)$ (with $N=2n$), obtaining tensors in 
\[
\left( \frac{n(n+1)}{2}\oplus \frac{\overline{n(n+1)}}{2}\right)\oplus \left( n\oplus \bar{n} \right).
\]
In particular, gauging $SU(5)\times U(1)$ ($n=5$), we get tensors in $(\mathbf{15}\oplus \mathbf{\overline{15}})\oplus(\mathbf{5}\oplus\mathbf{\bar{5}})$.

\item
$J^{\mathbb{C}}_{(1,N)}$\\
If we gauge $SU(N)\times U(1)\subset SU(N,1)$, we get tensors in the $\mathbf{N} \oplus \mathbf{\overline{N}}.$  In particular, taking $N=5$, we get $\mathbf{5}\oplus \mathbf{\bar{5}}$ tensors.  More generally, consider gauging $SU(5)\times U(1)\subset SU(5n,1)$; we get $n$ sets of tensors in the $(\mathbf{5}\oplus\mathbf{5})$ of $SU(5)$.

Next consider $N=27$ and gauge $E_{6}\times U(1)\subset SU(27)\times U(1)\subset SU(27,1)$.  This yields 704 fields outside of the adjoint representation.    
\item
$J^{\mathbb{H}}_{(1,N)}$\\
Gauging $SU(N,1)\subset USp(2N,2)$, we get tensors in $\frac{N(N+1)}{2}\oplus \frac{\overline{N(N+1)}}{2}$  Choosing $N=5$, we get $\mathbf{15}\oplus\mathbf{\overline{15}}$.  Under $SU(5)\times U(1)$, this becomes $(\mathbf{\bar{5}}\oplus\mathbf{10})\oplus(\mathbf{5}\oplus\mathbf{\overline{10}})$.    
\item
$J^{\mathbb{O}}_{(1,2)}$\\
The symmetry group of the Lagrangian is too small to gauge a GUT group.
\vspace{4mm}\\
We summarize the theories with reasonable numbers of tensor couplings in table 5.  They are all of $SU(5)\times U(1)$ type gauging.
                    
\end{itemize}
\vspace{4mm}
\begin{center}
\begin{tabular}{|c|c|c|} \hline
J.A. & $\mbox{dim}(\mathcal{M}_{R})$ & Tensor $K$-reps  \\ \hline
$J^{\mathbb{R}}_{(1,10)}$ & $64$ & $(\mathbf{15}\oplus \mathbf{\overline{15}})\oplus(\mathbf{5}\oplus\mathbf{\bar{5}})$ \\
$J^{\mathbb{C}}_{(1,5n)}$ & $5n(5n+2)-1$ & $n(\mathbf{5}\oplus \mathbf{\bar{5}})$ \\
$J^{\mathbb{H}}_{(1,5)}$ & $64$ & $(\mathbf{\bar{5}}\oplus\mathbf{10})\oplus(\mathbf{5}\oplus\mathbf{\overline{10}})$ \\

\hline
\end{tabular}
\end{center}
\vspace{2mm}
Table 5: Summary of theories admitting $SU(5)\times U(1)$ gauging with tensor couplings (and with smallest field content).  

\subsection{Summary and discussion: YMESGTs coupled to tensors}

Within the class of theories discussed in this paper, the only ones admitting reasonable numbers of tensor multiplets of interest in GUTs are the those based on Lorentzian Jordan algebras discussed in the previous section.  Table 5 lists the theories for the case of $SU(5)\times U(1)$ gauge group (they were discussed in~\cite{GZ:03apr}).  Generically, the gauging GUT groups starting from a given MESGT is associated with large numbers of tensors and spectator vector fields.  We have not systematically considered $E_{6}\times U(1)$ gauging with tensor multiplets, though it appears that these theories also have large numbers of tensors and singlets.  A large number of unwanted tensor multiplets can be troublesome in orbifold-GUT models since one cannot get rid of all of the field content in these multiplets via orbifold boundary conditions (this is shown in~\cite{SM2}). 

There are other families of tensor couplings with non-homogeneous scalar manifolds that we have not discussed in this paper.  Although the geometry of these scalar manifolds is not understood, an algebraic discussion of such theories can be found in~\cite{EGZ:01aug}.  

Non-trivially charged tensor multiplets offer a possible way to introduce scalar fields in non-trivial gauge group representations in a more economical way; for every hypermultiplet introduced, there are four real scalars, whereas for every tensor muliplet introduced, there is only one real scalar.  Such a $5D$ GUT model is considered in~\cite{DGKL:01nov}.  The authors consider a field content that consists of gauge multiplets for $SU(5)$, $10$ tensor multiplets (in the $\mathbf{5}\oplus\mathbf{\bar{5}}$ of $SU(5)$), a spectator vector multiplet, and a graviton multiplet.  The theory is not obtained from an explicit scalar manifold, though and appears to be obtained by decomposition of $SU(5)\subset G$, with $G=SU(6)$ or a non-compact form thereof.  However, one must recall that, if tensor mulitplets non-trivially charged under a compact gauge group $K$, there must be at least one abelian factor (corresponding to an isometry of the real scalar manifold).  \vspace{2mm}\\
\textbf{Example of gauging with tensor couplings}

Within the framework of supergravity that we have reviewed throughout, the above model can be obtained from the theory based on Lorentzian Jordan algebra $J^{\mathbb{C}}_{(1,5)}$.  Gauging $SU(5)\times U(1)\subset SU(5,1)$ yields a theory with the $35$ vector fields decomposing into $\mathbf{1}\oplus\mathbf{5}\oplus\mathbf{\bar{5}}\oplus\mathbf{24}$ under $U(5)$.  The $\mathbf{5}\oplus\mathbf{\bar{5}}$ vectors must be dualized to tensor fields, and the singlet gauging the $U(1)$ is the graviphoton.  Under particular boundary conditions of the theory, one can obtain massless chiral multiplets in the $\mathbf{5}\oplus\mathbf{\bar{5}}$ along with their CPT conjugates, which can therefore potentially serve as a Higgs sector.  Otherwise, these will lead to massive vector multiplets in the $\mathbf{5}\oplus\mathbf{\bar{5}}$.  More details of this will be shown in~\cite{SM2}, which decribes the options for parity assignments in $5D$ $\mathcal{N}=2$ orbifold supergravity theories.  \textit{Note that the graviphoton is required in the gauging}.  This theory comes from the gauging of a \textbf{unified} MESGT, where all of the fields had been connected by a continuous global symmetry; the gauging of $SU(5)\times U(1)$ then disconnects the gravity and vector supermultiplets.      

Obtaining a simple unifying group appears difficult starting from a theory with $K_{simple}\times K_{abelian}$ and charged tensors.  One could be satisfied with a partially unified group like the ``flipped SU(5)" model with $SU(5)\times U(1)$ gauge group~\cite{flipped}.  But suppose we wish to embed the above theory into one with a simple compact gauge group.  There will be tensors at each stage of the embedding, requiring a U(1) factor, until the tensor reps lie in the adjoint of the simple group.  Starting from $SU(5)\times U(1)$, that group is $E_{8}$.  However, starting with such a gauge group, there must be a mechanism for breaking it down the line to $SU(5)\times U(1)$; a Higgs mechanism will yield massive vectors, not tensors.  So we cannot embed the tensor coupled theory in this way.  Anyway, we don't know of a real scalar manifold admiting an $E_{8}$ gauging unless we take the theory defined by $C_{ijk}=0$.  But tensor couplings require non-trivial C-tensor components (see appendix C.2).  It could be that tensor multiplets have a natural home in higher dimensional unification scenarios.  

There is an alternative scenario in $5D$ that allows tensor coupings with simple groups: \textit{non-compact} gaugings.  This is the topic of the next section.  

\section{Non-compact gaugings and \textbf{unified} YMESGTs}

In contrast to their rigid limits, $5D$ $\mathcal{N}=2$ supergravity theories coupled to vector multiplets 
admit non-compact non-abelian gauge groups while remaining unitary.  Such gaugings have been considered since the 1980's~\cite{GST:83, GZ:03apr}.  The ground states of these theories preserve at most the maximal compact subgroup as a symmetry group, with the non-compact gauge multiplets becoming $5D$ BPS massive: 
\[
\{A^{M}_{\mu},\lambda^{m\,i}\},
\]      
where $M$ is the index for the non-compact generators of the gauge group (and $m$ is the tangent space index for the scalar manifold coresponding to the directions of the non-compact gauged isometries).  \vspace{2mm}\\  
\textbf{Example of Higgs sector from non-compact gauging}

Consider the infinite family of YMESGTs based on the Lorentzian Jordan algebras $J^{\mathbb{C}}_{(1,N)}$ with gauge group $SU(N,1)$.  This family of YMESGTs are known as \textbf{unified} in the sense that there is a continuous (and in this case local) symmetry relating every field in the theory; see the schematic in section 6.3, where horizontal arrows represent (local) supersymmetry transformations, and vertical arrows now represent the action of a \textit{simple} gauge group involving \textit{all} the vector fields of the theory (For more on \textbf{unified} YMESGTs, see~\cite{GST:83, GZ:03apr, GMZ:05}.)  The ground state of these theories preserves at most an $SU(N)\times U(1)$ symmetry group, while the $2N$ non-compact gauge fields transforming in the $\mathbf{N}\oplus \mathbf{\bar{N}}$ become BPS massive.   

In particular, we may consider the $N=5$ case: we have an $SU(5,1)$ gauge group with $35$ gauge fields.  The ground state of the theory can have at most $SU(5)\times U(1)$ gauge group with the remaining $10$ vector multiplets in the $\mathbf{5}\oplus \mathbf{\bar{5}}$ becoming BPS massive.  This could be a unified theory into which flipped $SU(5)$ is embedded, since $U(5)$ is not simple, but $SU(5,1)$ is.  Note that the scalars in the non-compact gauge multiplets are eaten by the vector partners, and so we cannot obtain $5D$ Higgs scalars from these theories.  Upon dimensional reduction, a $5D$ $\mathcal{N}=2$ BPS vector multiplet yields a $4D$ $\mathcal{N}=2$ BPS vector multiplet: 
\[\{A^{M}_{\mu},\lambda^{m\,i},\,A^{M}\},\]
where the real scalar fields $A^{M}$ come from the reduction of $5D$ vectors.  Truncating to $\mathcal{N}=1$, we would expect a massive $4D$ $\mathcal{N}=1$ vector multiplet; we would not obtain $4D$ massless Higgs scalars.  However, just as in the case of tensor multiplets on the orbifold, one can assign parities appropriately so that instead of a massive vector multiplet, we are left with massless $4D$ $\mathcal{N}=1$ chiral multiplets in the $\mathbf{5}\oplus\mathbf{\bar{5}}$ along with their CPT conjugate multiplets.  One can use orbifold parity conditions to obtain a subgroup of $SU(5)\times U(1)$ and/or use a vev on the boundaries to perform the breaking, as is usually done in orbifold models.  \vspace{2mm}\\
\textbf{Example of non-compact gauging with tensors}

One can also now have tensor multiplets charged under a simple gauge group.  For example, one can gauge $SU(N,1)$ in the theory based on the Lorentzian Jordan algebra $J^{\mathbb{H}}_{(1,N)}$, with charged tensor multiplets in the $\mathbf{\frac{(N+1)}{2}}\oplus \mathbf{\frac{\overline{(N+1)}}{2}}$ (see section 6.3).  The theory is a \textbf{unified} YMESGT coupled to tensors.  The original MESGT is a \textbf{unified} theory, but when gauged, the tensor sector is ``cut off".  In particular, consider the case of $N=5$; we get the \textbf{unified} gauge group $SU(5,1)$ again, but now coupled to tensors in the $\mathbf{15}\oplus\mathbf{\overline{15}}$.  The ground state has at most $SU(5)\times U(1)$ gauge group, with $\mathbf{5}\oplus\mathbf{\bar{5}}$ BPS massive vector multiplets, and tensor multiplets in the $(\mathbf{\bar{5}}\oplus\mathbf{10})\oplus(\mathbf{5}\oplus\mathbf{\overline{10}})$.  This theory can then be orbifolded~\cite{SM2}.        

Though one may obtain non-compact gauged supergravity theories from ``compactification" of M- or string theory on non-compact hyperboloidal manifolds~\cite{CGP:04jan}, it is not clear that this is the only way to obtain such gaugings from a more fundamental theory in higher dimensions.   

\section{Conclusion}

Using the results of ~\cite{GST:84, GZ:03apr}, we have pointed out that, within the classification of homogeneous quaternionic scalar manifolds, one cannot choose any pairing of gauge group and hypermultiplet representations one likes.  However, we have shown how orbifold models considered in the literature can be constructed in supergravity via these manifolds (without the introduction of boundary localized fields).    
\begin{itemize}
\item
Any number $n$ of $5D$ Higgs hypermultiplets in the $\mathbf{5}$ or $\mathbf{10}$ can be embedded in an $SU(5)$ or $SO(10)$ (resp.) gauge theory by coupling the relevant homogeneous scalar manifolds in tables 4 and 3 (resp.).
\item
If a single generation of matter in the $\mathbf{5}\oplus\mathbf{10}$ of an $SU(5)$ theory is desired, an additional singlet must be coupled (see table 4).  The theory is characterized by the coupling of the scalar manifold (see table 1)
\[
\mathcal{M}_{R}\times \frac{E_{7}}{\overline{SO(12)}\times SU(2)}.
\] 
But this theory admits an $SO(10)$ gauging under which the matter forms the $\mathbf{16}$ (table 3).
\item
If $n$ generations of bulk matter \textit{and} Higgs multiplets in the $$(\mathbf{\bar{5}}\oplus\mathbf{10})\oplus(\mathbf{5}\oplus\mathbf{\bar{5}})$$ of the gauge group $SU(5)$ are desired, two additional singlets must be added (see table 4).  This corresponds to the coupling of the scalar manifold (see table 1) 
\[
\mathcal{M}_{R}\times \frac{SU(27n,2)}{SU(27n)\times SU(2)\times U(1)}.
\]
But this theory admits an $E_{6}$ gauging under which each generation of fields form the $\mathbf{27}$ (table 2).  
\end{itemize}
Since there are no known quaternionic scalar manifolds with compact $E_{8}$ isometries, there isn't a unification of generations into an irreducible representation within the framework of $5D$ supergravity.  However, multiple generations arise in Calabi-Yau compactifications of string or M-theory, and are therefore expected to appear in the supergravity approximations in four and five dimensions, respectively.  

In all of the above cases, $\mathcal{M}_{R}$ is whatever scalar manifold one has chosen, with the following constraints:
\vspace{2mm}\\
\textbf{(i)} The isometry group must admit the gauge group as a subgroup, and \\
\textbf{(ii)} If the decomposition of the $n_{V}+1$ vector fields under the compact gauge group has non-singlet (not including the adjoint) representations, the corresponding vectors must be dualized to tensors and at least one abelian isometry must be gauged.\vspace{1mm}

For minimality, one may take the space described in section 4 in which all of the vector fields gauge the group of interest, except the graviphoton, which is a spectator in the compact gaugings of scalar manifold isometries.  However, it should be emphasized that this choice is not motivated by any other compelling phenomenological reason.    

As an alternative scenario to hypermultiplets in the bulk, we may attempt to put the Higgs scalars or matter fields in tensor multiplets; however, gauging a compact subgroup of $Iso(\mathcal{M}_{R})$ with charged tensors requires the group to have at least one abelian factor.  For example, we may consider the $\mathbf{5}\oplus\mathbf{\bar{5}}$ tensor multiplets of the $SU(5)\times U(1)$ gauge theory based on the Lorentzian Jordan algebra $J^{\mathbb{C}}_{(1,5)}$.  This is a more economical approach, in the sense that one uses only 10 scalars as opposed to the 40 (in this example).  Even though theories with tensor couplings do not involve a simple gauge group, there is the partial unification model based on $SU(5)\times U(1)$ (``flipped $SU(5)$").  Furthermore, a generation of matter can sit in a bulk $(\mathbf{\bar{5}}\oplus\mathbf{10})\oplus(\mathbf{5}\oplus\mathbf{\overline{10}})$ by coupling the real scalar manifold based on the Lorentzian Jordan algebra $J^{\mathbb{H}}_{(1,5)}$.  We will thus be left with a non-chiral theory.           

Non-compact gauge groups in $5D$ supergravity are another novel way to get $4D$ massless chiral multiplets in interesting representations of the gauge group.  We have mentioned the example of a \textbf{unified} $SU(5,1)$ YMESGT based on the Lorentzian Jordan algebra $J^{\mathbb{C}}_{(1,5)}$, though there is an infinite family of such non-compact gaugings; the theories based on $J^{C}_{(1,N)}$ admit \textbf{unified} $SU(N,1)$ YMESGTS.  \textbf{Unified} here, is in the sense that all fields of the YMESGT are connected by a combination of supersymmetry and gauge transformations; therefore, this is in some sense in between the ideas of Grand Unification and gauge/gravity coupling unification.        

\section{Future Directions}

In a subsequent paper, we shall discuss more general orbifold supergravity theories, with focus on the classes of allowed $\mathbb{Z}_{2}$ parity assignments.  We will pay particular attention to implications in orbifold-GUTs, though, and much of it will be complementary to the discussions throughout the current paper.  Although many of the parity assignments have been used by a number of other authors over the years, we want to clear up the options in the framework of $\mathcal{N}=2$ $5D$ supergravity, and discuss the resulting classes of theories one can obtain based on the choices made.  We argue that these classes of theories are in correspondence with \textit{particular} classes of compactifications of M-theory on singular $7$-spaces of $G_{2}$ holonomy (those that admit an intermediate range of energies where the universe appears five dimensional); see~\cite{ES:96nov}, e.g., for related discussion.\vspace{4mm}\\
\textit{\textbf{Acknowledgements}}\\
The author would like to thank Murat G\"{u}naydin for proposal of the investigation, overall guidance, and helpful technical and editorial comments regarding this paper; Marco Zagermann for helpful comments regarding the manuscript; and finally, special thanks to Jonathan Bagger for enlightening conversation.\vspace{4mm}\\
\textbf{\Large{Appendices}}
\appendix
\section{$5D$ $\mathcal{N}=2$ Maxwell-Einstein Supergravity~\cite{GST:84}}  
An $\mathcal{N}=2$ $5D$ Maxwell-Einstein supergravity theory (MESGT) consists of a gravity supermutiplet $\{e^{m}_{\mu}, \Psi^{i}_{\mu}, A^{0}_{\mu}\}$ and $n_{V}$ vector supermultiplets $\{A^{i}_{\mu},\lambda^{i\,a},\phi^{x}\}$.  The total field content is therefore
$$\{e^{m}_{\mu}, \Psi^{i}_{\mu}, A^{I}_{\mu}, \lambda^{i\,a}, \phi^{x}\},$$
where $I=(0,1,...,n_{V})$ labels the graviphoton and vector fields from the $n_{V}$ vector multiplets; $i=(1,2)$ is an $SU(2)_{R}$ index; and
$a=(1,...,n_{V})$ and $x=(1,...,n_{V})$ label the
fermions and scalars from the $n_{V}$ vector
multiplets.  The scalar fields parametrize an $n_{V}$-manifold $\mathcal{M}_{R}$, so the indices $a,b,\ldots$ and
$x,y,\ldots$ may also be
viewed as
flat and curved indices of $\mathcal{M}_{R}$, respectively. 

The $5D$ bosonic Lagrangian is
\begin{equation}
\begin{split}
e^{-1}\mathcal{L}_{bos}= & -\frac{1}{2}R -
\frac{1}{4}\dot{a}_{IJ}(\phi)F^{I}_{\mu\nu}F^{J\,\mu\nu}
-\frac{1}{2}g_{xy}(\phi)(\partial_{\mu}\phi^{x})(\partial^{\mu}\phi^{y})\\
&
+\frac{e^{-1}}{6\sqrt{6}}C_{IJK}\epsilon^{\mu\nu\rho\sigma\lambda}
F^{I}_{\mu\nu}F^{J}_{\rho\sigma}A^{K}_{\lambda},
\end{split}
\end{equation}
where we are using a spacetime metric with signature $(-++++).$  The $A^{I}_{\mu}$ are abelian vector
fields: $F^{I}_{\mu\nu}=\partial_{[\mu}A^{I}_{\nu]}$.  The quantity $\hat{e}$ is the determinant of
the f\"{u}nfbein. The rank
three tensor $C_{IJK}$ is constant and symmetric.  The tensors $\dot{a}_{IJ}(\phi)$ and $g_{xy}(\phi)$ are symmetric and scalar dependent.   

We will now describe the target space geometry of the theory.  Introducing $(n_{V}+1)$ parameters $\xi^{I}(\phi)$ depending on the scalar fields, the tensor
$C_{IJK}$ appearing in the Lagrangian can be
associated with a cubic 
polynomial
$$\mathcal{V} (\xi)=C_{IJK}\xi^{I}\xi^{J}\xi^{K}.$$
This polynomial defines a symmetric tensor 
$$a_{IJ}(\xi)=-\frac{1}{3}\frac{\partial}{\partial
\xi^{I}}\frac{\partial}{\partial \xi^{J}} \ln\mathcal{V} (\xi).$$  
The parameters $\xi^I$ can be interpreted as coordinate functions for an   
$(n_{V}+1)$-manifold, which we call the ambient space.  
The tensor $a_{IJ}$, which may have indefinite signature, defines a metric on
this space.  However, the coordinates are restricted via
$\mathcal{V} (\xi)>0$ so that the metric is positive definite, which means that
the manifold is Riemannian.  The equation $\mathcal{V} (\xi)=k$ ($k\in \mathbb{R}$) defines a
family of real
hypersurfaces, and in particular $$\mathcal{V} (\xi)=1$$ defines a real $n_{V}$-manifold corresponding to the scalar manifold $\mathcal{M}_{R}$.  The functions $h^{I}$ and $h^{I}_{x}$ that appear in the full Lagrangian (they appear explicitly in fermionic terms) are directly proportional to $\xi^{I}|_{\mathcal{V}=1}$ and $\xi^{I}_{,x}|_{\mathcal{V}=1}$, respectively; the $h^{I}$ are essentially embedding coordinates of $\mathcal{M}_{R}$ in the ambient space.    

The tensor $\dot{a}_{IJ}$ appearing in
the kinetic term for the abelian fieldstrengths is the restriction of the
``ambient space" metric to $\mathcal{M}_{R}$:
$$\dot{a}_{IJ}=a_{IJ}|_{\mathcal{V} =1}.$$
The tensor $g_{xy}$ in the kinetic term for the scalar fields (the metric of the scalar
manifold) is the pullback of the restricted ambient space metric to $\mathcal{M}_{R}$:
$$g_{xy}=\frac{3}{2}\dot{a}_{IJ}h^{I}_{,x}h^{J}_{,y}.$$
Both of these tensors contracting the fieldstrengths and scalars in the kinetic terms are positive definite (due to the constraint
$\mathcal{V}>0$ we imposed), and therefore the theories are free of ghosts.  

Since the $C_{IJK}$ tensor completely determines a MESGT, the global symmetry group of the Lagrangian is given by the symmetry group, $G$, of this tensor, along with automorphisms of the superalgebra: $G\times SU(2)_{R}$.  The $G$-action on the ambient space can be represented by $\xi^{I}\rightarrow M^{I}_{J}\xi^{J}$ (and similarly for the vector fields), with the $M^{I}_{J}$ satisfying the condition 
\[
M^{I'}_{(I}C_{JK)I'}=0,
\]
where parentheses denote symmetric permutations.  Since $G$ consists of symmetries of the full Lagrangian, they are symmetries of the scalar sector in particular, and therefore isometries of the scalar manifold $\mathcal{M}_{R}$: $G\subset Iso(\mathcal{M}_{R})$.  The full Lagrangian, however, is not necessarily invariant under the full group $Iso(\mathcal{M}_{R})$.
   
\section{Coupling hypermultiplets to MESGTs~\cite{BW:83}}
We now consider the addition of $n_{H}$ hypermultiplets $\{\zeta^{A},q^{X}\}$, where $A=1,\ldots,2n_{H}$ and $X=1,\ldots,4n_{H}$.  The scalars $q^{X}$ parametrize an $n_{H}$-dimensional quaternionic manifold $\mathcal{M}_{Q}$ ($4n_{H}$ real dimensions), where $X$ and $A$ are curved and tangent space indices, respectively; in addition, there are $USp(2)\simeq SU(2)$ flat indices of the quaternionic manifold, since the tangent space group is $USp(2n_{H})\times USp(2)$.  

The covariant derivative of the ungauged theory is a linear combination of the flat spacetime connection, the spacetime spin connection, and $USp(2n_{H})\times USp(2)$ connections.  

\section{Gauging of $\mathcal{N}=2$ $5D$ supergravity}

The vector fields in supergravity theories may be elevated to gauge fields.  The gauge symmetries come from elevating global symmetries of the Lagrangian to local symmetries.  As discussed in appendix A, the global symmetry group of a MESGT is $G\times SU(2)_{R}$, where $G\subset Iso(\mathcal{M}_{R})$ and $SU(2)_{R}$ is the automorphism group of the superalgebra.  We use the terminology of~\cite{GST:85} when the following groups, $K$, are gauged:
\[
\begin{array}{cc}
K\subset G & \mbox{Yang-Mills-Einstein supergravity theory (YMESGT)}\\
K\subset SU(2)_{R} & \mbox{Gauged Maxwell-Einstein supergravity theory}\\
K\subset G\times SU(2)_{R} & \mbox{Gauged YMESGT}.
\end{array}  
\]
The extension to MESGTs coupled to hypermultiplets goes the same way.  Now, the isometry group of the total scalar manifold is $Iso(\mathcal{M})\equiv Iso(\mathcal{M}_{R})\times Iso(\mathcal{M}_{Q})$.  The global symmetry group of the Lagrangian is $G\times SU(2)_{R}$, where now $G\subset Iso(\mathcal{M})$.   

The bulk spacetime (preserving supersymmetry) that one will get depends on the choice of scenario:\\  
A pure YMESGT does not have scalar potential, so that the bulk spacetime will be flat.\\  
A YMESGT coupled to tensors does have a scalar potential, which admits supersymmetric Minkowski vacua.\\
Coupling to hypermultiplets introduces a scalar potential also admitting supersymmetric AdS vacua.\\
We have not discussed the case of a gauged YMESGT (where a subgroup of $SU(2)_{R}$ is gauged), which admits AdS or flat supersymmetric vacua, depending on the linear combination of vector fields used to gauge this factor (coupling to tensors then results in novel supersymmetric vacua)~\cite{gauged, GST:85, GZ:99dec, GZ:00}.   

\subsection{Simple class of YMESGTs without tensor- or hyper-multiplets~\cite{GST:84}}

We begin with a $5D$ $\mathcal{N}=2$ MESGT, which consists of an $\mathcal{N}=2$ gravity supermultiplet coupled to $n_{V}$ vector supermultiplets.  As discussed in appendix A, the theory is completely determined by a rank three symmetric tensor $C_{IJK}$, which can be put in the canonical form
\begin{equation}
C_{000}=1,\;\;C_{0ij}=-\frac{1}{2}\delta_{ij},\;\;C_{00i}=0,\;\;C_{ijk}=\mbox{arbitrary},
\end{equation}
where $i,j,k=1\ldots n_{V}$.
The final expression above represents the existence of a myriad of scalar manifolds for a given value of $n_{V}$.  The scalar manifold, $\mathcal{M}_{R}$, is a real Riemannian $n_{V}$-manifold.  The (possibly trivial) isometry group is denoted $Iso(\mathcal{M_{R}})$.  The global symmetry group of the Lagrangian, $G$, is the subgroup that leaves the $C$-tensor invariant.  

There are many YMESGTs one can consider, characterized by various choices of $C_{IJK}$, and the gauging generally requires a number of spectator vector fields and/or the presence of charged tensors (the latter requires an abelian gauge group factor; see section 3).  But here we will be interested in particular compact gaugings, with minimal additional field content to make matters simple.  There are then two forms of $C_{IJK}$ we are interested in.  
\vspace{2mm}\\
\textbf{(1)}  Consider the simple theory with $n_{V}$ vector multiplets and 
\[C_{ijk}=0.\]
The scalar manifold for this choice of theories is in general non-homogeneous, and the Lagrangian is invariant under the maximum possible group $G=SO(n_{V})$ (as is clear from the form of $C_{ijk}$).  The vector fields decompose as $\mathbf{n_{V}}\oplus \mathbf{1}$ under $G$.  Thus, the vector fields other than the graviphoton transform in the fundamental representation of $SO(n_{V})$; the graviphoton is a spectator vector field.  \\
\textit{Remark}: any choice of $C_{ijk}\neq 0$ breaks the global symmetry group of the Lagrangian to a subgroup of $SO(n_{V}),$ and the $n_{V}$ vector fields will no longer necessarily form an irrep of this new symmetry group.  

The adjoint representation of any compact group $K$ can always be embedded in the fundamental representation of $SO(n_{V})$ with $n_{V}\geq \mbox{dim}(K)$;  the $n_{V}-\mbox{dim}(K)$ vectors join the graviphoton as spectators.  

It follows that, in our case, the adjoint representation of $K$ can be exactly embedded into the $\mathbf{n_{V}}$ of $SO(n_{V})$ without additional fields (i.e., $n_{V}=\mbox{dim}(K)$.  In particular, consider $n_{V}=24$ and $K=SU(5)$; $n_{V}=45$ and $K=SO(10)$; or $n_{V}=78$ and $K=E_{6}$.  In this way, we can obtain an $SU(5)$, $SO(10)$, or $E_{6}$ YMESGT with singlet graviphoton.  Of course, one may consider other compact gaugings similarly.  All fields in the vector supermultiplets will be in the adjoint representation of the gauge group, while all fields in the graviton supermultiplet are gauge singlets.
\vspace{2mm}\\ 
\textbf{(2)}  This is not the end of the story, though.  One may split the index $i=(a,\alpha)$, and take $C_{ijk}$ to be
\[
 C_{abc}=b d_{abc};\;\;\;\;C_{\alpha\beta\gamma}=0,
\]
where $d_{abc}$ are the $d$-symbols of $SU(n)\subset SO(n_{V})$; and $b\geq 0$ is a real parameter.  The group action preserving the $C$-tensor is consequently reduced to a subgroup $SU(n)\subset SO(n_{V})$.  Now, $K\subset SU(n)$ can be gauged, with the remaining $n_{V}-\mbox{dim}(K)$ vector fields outside the adjoint representation being spectators.  Again, if we are interested in minimal field content, then we demand that $n_{V}=\mbox{dim}(K)$, which means that we must restrict out attention to $SU(n)$ gaugings with $\mbox{dim}[SU(n)]=n_{V}$.  There are then no vector fields with $\alpha$ indices (i.e., no singlets).   

\textit{Remark}:  In contrast to the case of $b=0$, for $b\neq 0$ we have a single parameter family of theories.  The theories are of the same form, since the $C$-tensor determines couplings in the theory, but there is a single adjustable parameter affecting the strength of those couplings.  
 
\subsection{YMESGTs coupled to tensor multiplets~\cite{GZ:99dec}}

When a MESGT with $n_{V}$ abelian vector multiplets is gauged, the symmetry group of the Lagrangian is broken to the gauge group $K\subset G$.  The $n_{V}+1$ vector fields decompose into $K$-reps 
\[ \mathbf{n_{V}+1}=\mathbf{adj(K)}\oplus \mathbf{non\mbox{-}singlets(K)}\oplus \mathbf{singlets(K)}. \]   
The $K$-transformations of the non-singlet vector fields would break the abelian symmetry associated with each vector, thereby making the corresponding fields massive, which in turn would break supersymmetry.  Thus, gauging requires the non-singlet vector fields to be dualized to anti-symmetric tensor fields satisfying a field equation that serves as a ``self-duality" constraint (thus keeping the degrees of freedom the same):
\[ B^{M}_{\mu\nu}=c^{M}_{N}\,\epsilon^{\;\;\;\rho\sigma\lambda}_{\mu\nu}\,\partial_{[\rho}B^{N}_{\sigma\lambda]}+\cdots, \]   
where $c^{M}_{N}$ has dimensions of inverse mass; square brackets denote anti-symmetric permutations; and ellipses denote terms involving other fields.  A tensor field does not require an abelian invariance to remain massless.  

Let us split the vector index $\tilde{I}$ into $\tilde{I}=(I,M)$, which is a singlet/gauge index and non-singlet index, respectively.  To be consistent with the gauge symmetry, the components of the $C$-tensor are constrained to be:\footnote{We are assuming a compact or non-semisimple gauge group $K$; see~\cite{Bergshoeff} for more general couplings where $C_{MIJ}\neq 0$.} 
\[
\begin{split}
C_{IMN}=\frac{\sqrt{6}}{2}\Omega_{NP}\Lambda^{P}_{IM}\\
C_{MNP}=0;\;\;C_{MIJ}=0,
\end{split}
\]  
where $\Omega_{NP}$ is antisymmetric and $\Lambda^{P}_{IM}$ are symplectic $K$-representation matrices appearing in the $K$-transformation of the tensor fields: $\delta_{\alpha} B^{M}_{\mu\nu}=\alpha^{I}\Lambda^{M}_{IN}B^{N}_{\mu\nu}.$
Furthermore, $C_{IJK}$ must be a rank-three symmetric $K$-invariant tensor.

The new terms in the bosonic Lagrangian are 
\[
\begin{split}
\mathcal{L}=-\frac{e}{4}\dot{a}_{\tilde{I}\tilde{J}}&\mathcal{H}^{\tilde{I}}_{\mu\nu}\mathcal{H}^{\tilde{J}\;\mu\nu}\\
&+\frac{1}{4g}\epsilon^{\mu\nu\rho\sigma\lambda}\Omega_{MN}B^{M}_{\mu\nu}\mathcal{D}_{\rho}B^{N}_{\sigma\lambda} 
-eg^{2}P^{(T)},
\end{split}
\]
where 
\begin{gather}
\mathcal{H}^{\tilde{I}}_{\mu\nu}:= (\mathcal{F}^{I}_{\mu\nu},\,B^{M}_{\mu\nu})
\mathcal{D}_{\mu}B^{M}_{\nu\rho}\equiv \nabla_{\mu}B^{M}_{\nu\rho} + gA^{I}_{\mu}\Lambda^{M}_{IN}B^{N}_{\nu\rho}.\nonumber
\end{gather}

The presence of non-trivially charged tensors introduces a scalar potential $P^{(T)}$ that was not present in the case of pure YMESGTs.  It is given by
\begin{gather}
P^{(T)}=2W^{a}W^{a}\\
W^{a}=-\frac{\sqrt{6}}{8}h^{a}_{M}\Omega^{MN}h_{N},
\end{gather}
where the $h^{M}$ are functions of the scalar fields in the tensor multiplets (see appendix A).


\begin{thebibliography}{10}

\bibitem{sin2:91}
U.~Amaldi, W.~de Boer and H.~Furstenau,
``Comparison of grand unified theories with electroweak and strong coupling
constants measured at LEP,''
Phys.\ Lett.\ B {\bf 260}, 447 (1991);
C.~Giunti, C.~W.~Kim and U.~W.~Lee,
``Running coupling constants and grand unification models,''
Mod.\ Phys.\ Lett.\ A {\bf 6}, 1745 (1991);
J.~R.~Ellis, S.~Kelley and D.~V.~Nanopoulos,
``Probing The Desert Using Gauge Coupling Unification,''
Phys.\ Lett.\ B {\bf 260}, 131 (1991);
P.~Langacker and M.~x.~Luo,
``Implications of precision electroweak experiments for M(t), rho(0),
sin**2-Theta(W) and grand unification,''
Phys.\ Rev.\ D {\bf 44}, 817 (1991).

\bibitem{KD:96feb}
K.~R.~Dienes,
``String Theory and the Path to Unification: A Review of Recent Developments,''
Phys.\ Rept.\  {\bf 287}, 447 (1997)
[arXiv:hep-th/9602045].

\bibitem{NS:97feb}
H.~P.~Nilles and S.~Stieberger,
``String unification, universal one-loop corrections and strongly coupled
heterotic string theory,''
Nucl.\ Phys.\ B {\bf 499}, 3 (1997)
[arXiv:hep-th/9702110].


\bibitem{NS:95oct}
H.~P.~Nilles and S.~Stieberger,
``How to Reach the Correct $sin^{2}\theta_W$ and $\alpha_S$ in String Theory,''
Phys.\ Lett.\ B {\bf 367}, 126 (1996)
[arXiv:hep-th/9510009].

\bibitem{HW}
P.~Horava and E.~Witten,
``Heterotic and type I string dynamics from eleven dimensions,''
Nucl.\ Phys.\ B {\bf 460}, 506 (1996)
[arXiv:hep-th/9510209];
E.~Witten,
``Strong Coupling Expansion Of Calabi-Yau Compactification,''
Nucl.\ Phys.\ B {\bf 471}, 135 (1996)
[arXiv:hep-th/9602070];
P.~Horava and E.~Witten,
``Eleven-Dimensional Supergravity on a Manifold with Boundary,''
Nucl.\ Phys.\ B {\bf 475}, 94 (1996)
[arXiv:hep-th/9603142].


\bibitem{KK}
G.~Chapline and N.~S.~Manton,
``The Geometrical Significance Of Certain Higgs Potentials: An Approach To
Grand Unification,''
Nucl.\ Phys.\ B {\bf 184}, 391 (1981);
P.~G.~O.~Freund,
``Grand Unification Near The Kaluza-Klein Scale,''
Phys.\ Lett.\ B {\bf 120}, 335 (1983).


\bibitem{guts}
Y.~Kawamura,
``Gauge symmetry reduction from the extra space S(1)/Z(2),''
Prog.\ Theor.\ Phys.\  {\bf 103}, 613 (2000)
[arXiv:hep-ph/9902423];
Y.~Kawamura,
``Triplet-doublet splitting, proton stability and extra dimension,''
Prog.\ Theor.\ Phys.\  {\bf 105}, 999 (2001)
[arXiv:hep-ph/0012125];
Y.~Kawamura,
``Split multiplets, coupling unification and extra dimension,''
Prog.\ Theor.\ Phys.\  {\bf 105}, 691 (2001)
[arXiv:hep-ph/0012352];
A.~B.~Kobakhidze,
``Proton stability in TeV-scale GUTs,''
Phys.\ Lett.\ B {\bf 514}, 131 (2001)
[arXiv:hep-ph/0102323];
G.~Altarelli and F.~Feruglio,
``SU(5) grand unification in extra dimensions and proton decay,''
Phys.\ Lett.\ B {\bf 511}, 257 (2001)
[arXiv:hep-ph/0102301];
A.~Hebecker and J.~March-Russell,
``A minimal S(1)/(Z(2) x Z'(2)) orbifold GUT,''
Nucl.\ Phys.\ B {\bf 613}, 3 (2001)
[arXiv:hep-ph/0106166];
M.~Kakizaki and M.~Yamaguchi,
``Splitting triplet and doublet in extra dimensions,''
Prog.\ Theor.\ Phys.\  {\bf 107}, 433 (2002)
[arXiv:hep-ph/0104103];
L.~J.~Hall, H.~Murayama and Y.~Nomura,
``Wilson lines and symmetry breaking on orbifolds,''
Nucl.\ Phys.\ B {\bf 645}, 85 (2002)
[arXiv:hep-th/0107245].

\bibitem{DM:01aug}
R.~Dermisek and A.~Mafi,
``SO(10) grand unification in five dimensions: Proton decay and the mu
problem,''
Phys.\ Rev.\ D {\bf 65}, 055002 (2002)
[arXiv:hep-ph/0108139].

\bibitem{HN:01mar}
L.~J.~Hall and Y.~Nomura,
``Gauge unification in higher dimensions,''
Phys.\ Rev.\ D {\bf 64}, 055003 (2001)
[arXiv:hep-ph/0103125];

\bibitem{DDG:98}
K.~R.~Dienes, E.~Dudas and T.~Gherghetta,
``Extra spacetime dimensions and unification,''
Phys.\ Lett.\ B {\bf 436}, 55 (1998)
[arXiv:hep-ph/9803466];
K.~R.~Dienes, E.~Dudas and T.~Gherghetta,
``Grand unification at intermediate mass scales through extra dimensions,''
Nucl.\ Phys.\ B {\bf 537}, 47 (1999)
[arXiv:hep-ph/9806292].

\bibitem{YL:03dec}
Y.~Lin,
``On orbifold theory and N = 2, D = 5 gauged supergravity,''
JHEP {\bf 0401}, 041 (2004)
[arXiv:hep-th/0312078].


\bibitem{ZGAZ:04jul}
F.~P.~Zen, B.~E.~Gunara, Arianto and H.~Zainuddin,
``On orbifold compactification of N = 2 supergravity in five dimensions,''
arXiv:hep-th/0407112.

\bibitem{DGKL:04feb}
G.~A.~Diamandis, B.~C.~Georgalas, P.~Kouroumalou and A.~B.~Lahanas,
``On the brane coupling of unified orbifolds with gauge interactions in the
bulk,''
arXiv:hep-th/0402228.

\bibitem{BW:83}
J.~Bagger and E.~Witten, ``Matter Couplings In N=2 Supergravity ,''
Nucl.\ Phys.\ B {\bf 222}, 1 (1983).

\bibitem{rigid}
L.~Andrianopoli, M.~Bertolini, A.~Ceresole, R.~D'Auria, S.~Ferrara, P.~Fre and T.~Magri,
``N = 2 supergravity and N = 2 super Yang-Mills theory on general scalar
manifolds: Symplectic covariance, gaugings and the momentum map,''
J.\ Geom.\ Phys.\  {\bf 23}, 111 (1997)
[arXiv:hep-th/9605032];
M.~Billo, F.~Denef, P.~Fre, I.~Pesando, W.~Troost, A.~Van Proeyen and D.~Zanon,
``The rigid limit in special Kaehler geometry: From K3-fibrations to  special
Riemann surfaces: A detailed case study,''
Class.\ Quant.\ Grav.\  {\bf 15}, 2083 (1998)
[arXiv:hep-th/9803228]

\bibitem{YN:01aug}
Y.~Nomura,
``Strongly coupled grand unification in higher dimensions,''
Phys.\ Rev.\ D {\bf 65}, 085036 (2002)
[arXiv:hep-ph/0108170].

\bibitem{KR:02dec}
H.~D.~Kim and S.~Raby,
``Unification in 5D SO(10),''
JHEP {\bf 0301}, 056 (2003)
[arXiv:hep-ph/0212348].

\bibitem{2higgs}
S.~Dimopoulos and H.~Georgi,
Nucl.\ Phys.\ B {\bf 193}, 150 (1981);
J.~R.~Ellis, S.~Kelley and D.~V.~Nanopoulos,
Phys.\ Lett.\ B {\bf 260}, 131 (1991);
U.~Amaldi, W.~de Boer and H.~Furstenau,
Phys.\ Lett.\ B {\bf 260}, 447 (1991);
C.~Giunti, C.~W.~Kim and U.~W.~Lee,
Mod.\ Phys.\ Lett.\ A {\bf 6}, 1745 (1991);
P.~Langacker and M.~x.~Luo,
Phys.\ Rev.\ D {\bf 44}, 817 (1991).

\bibitem{bailin}
D.~Bailin and A.~Love,
``Non-minimal Higgs content in standard-like models from D-branes at a Z(N) singularity,''
Phys.\ Lett.\ B {\bf 598}, 83 (2004);
D.~Bailin, G.~V.~Kraniotis and A.~Love,
``Supersymmetric standard models on D-branes,''
Phys.\ Lett.\ B {\bf 502}, 209 (2001)
[arXiv:hep-th/0011289];
D.~Bailin and A.~Love,
``Orbifold compactifications of string theory,''
Phys.\ Rept.\  {\bf 315}, 285 (1999);



\bibitem{NSW:01apr}
Y.~Nomura, D.~R.~Smith and N.~Weiner,
``GUT breaking on the brane,''
Nucl.\ Phys.\ B {\bf 613}, 147 (2001)
[arXiv:hep-ph/0104041].



\bibitem{DL:89aug}
D.~C.~Lewellen,
``Embedding Higher Level Kac-Moody Algebras In Heterotic String Models,''
Nucl.\ Phys.\ B {\bf 337}, 61 (1990).

\bibitem{FIQ:90mar}
A.~Font, L.~E.~Ibanez and F.~Quevedo,
``Higher Level Kac-Moody String Models And Their Phenomenological
Implications,''
Nucl.\ Phys.\ B {\bf 345}, 389 (1990).

\bibitem{DF:95aug}
D.~Finnell,
``Grand Unification with Three Generations in Free Fermionic String Models,''
Phys.\ Rev.\ D {\bf 53}, 5781 (1996)
[arXiv:hep-th/9508073].


\bibitem{GST:84}
M.~G\"{u}naydin, G.~Sierra and P.~K.~Townsend,
``The Geometry Of N=2 Maxwell-Einstein Supergravity And Jordan Algebras,''
Nucl.\ Phys.\ B {\bf 242}, 244 (1984).

\bibitem{GST:85}
M.~Gunaydin, G.~Sierra and P.~K.~Townsend,
``Gauging The D = 5 Maxwell-Einstein Supergravity Theories: More On Jordan
Algebras,''
Nucl.\ Phys.\ B {\bf 253}, 573 (1985).

\bibitem{EGZ:01aug}
J.~R.~Ellis, M.~Gunaydin and M.~Zagermann,
``Options for gauge groups in five-dimensional supergravity,''
JHEP {\bf 0111}, 024 (2001)
[arXiv:hep-th/0108094].

\bibitem{GRW:86}
M.~Gunaydin, L.~J.~Romans and N.~P.~Warner,
``Compact And Noncompact Gauged Supergravity Theories In Five-Dimensions,''
Nucl.\ Phys.\ B {\bf 272}, 598 (1986).

\bibitem{TPvN:84}
P.~K.~Townsend, K.~Pilch and P.~van Nieuwenhuizen,
``Selfduality In Odd Dimensions,''
Phys.\ Lett.\  {\bf 136B}, 38 (1984)
[Addendum-ibid.\  {\bf 137B}, 443 (1984)].

\bibitem{GZ:99dec}
M.~Gunaydin and M.~Zagermann,
Nucl.\ Phys.\ B {\bf 572}, 131 (2000)
[arXiv:hep-th/9912027].

\bibitem{CDA:00apr}
A.~Ceresole and G.~Dall'Agata,
``General matter coupled N = 2, D = 5 gauged supergravity,''
Nucl.\ Phys.\ B {\bf 585}, 143 (2000)
[arXiv:hep-th/0004111].

\bibitem{S:85}
G.~Sierra,
``N=2 Maxwell Matter Einstein Supergravities In D = 5, D = 4 And D = 3,''
Phys.\ Lett.\ B {\bf 157}, 379 (1985).

\bibitem{dWVP:95may}
B.~de Wit and A.~Van Proeyen,
``Isometries of special manifolds,''
arXiv:hep-th/9505097.

\bibitem{AS:97jun}
A.~Strominger,
``Loop corrections to the universal hypermultiplet,''
Phys.\ Lett.\ B {\bf 421}, 139 (1998)
[arXiv:hep-th/9706195].

\bibitem{GZ:03apr}
M.~G\"{u}naydin and M.~Zagermann,
``Unified Maxwell-Einstein and Yang-Mills-Einstein supergravity theories  in
five dimensions,''
JHEP {\bf 0307}, 023 (2003)
[arXiv:hep-th/0304109].

\bibitem{GST:83}
M.~Gunaydin, G.~Sierra and P.~K.~Townsend,
``Exceptional Supergravity Theories And The Magic Square,''
Phys.\ Lett.\ B {\bf 133}, 72 (1983).

\bibitem{GMZ:05}
M.~Gunaydin, S.~McReynolds, M.~Zagermann, Forthcoming paper.

\bibitem{PP:86}
D.~N.~Page and C.~N.~Pope,
``Einstein Metrics On Quaternionic Line Bundles,''
Class.\ Quant.\ Grav.\  {\bf 3}, 249 (1986).

\bibitem{KG:87}
K.~Galicki,
``New Matter Couplings In N=2 Supergravity,''
Nucl.\ Phys.\ B {\bf 289}, 573 (1987).

\bibitem{HP:87}
H.~Pedersen,
``Einstein metrics, spinning top motion, and monopoles,"
Math.\ Ann.\ {\bf 274}, 35 (1986)

\bibitem{AS:91}
A.~Swann,
``HyperK\"{a}hler and quaternionic K\"{a}hler geometry,"
Math.\ Ann.\ \textbf{289}, 421 (1991)


\bibitem{SM3}
S.~McReynolds,
``Five-dimensional Yang-Mills-Einstein Supergravity on Orbifolds: Symmetries, anomalies, and axions,"
Forthcoming. 

\bibitem{wilson}
L.~J.~Hall, H.~Murayama and Y.~Nomura,
  Nucl.\ Phys.\ B {\bf 645}, 85 (2002)
  [arXiv:hep-th/0107245];
N.~Haba, M.~Harada, Y.~Hosotani and Y.~Kawamura,
Nucl.\ Phys.\ B {\bf 657}, 169 (2003)
[Erratum-ibid.\ B {\bf 669}, 381 (2003)]
[arXiv:hep-ph/0212035];
J.~Bagger and M.~Redi,
Phys.\ Lett.\ B {\bf 582}, 117 (2004)
[arXiv:hep-th/0310086].

\bibitem{trinification}
S.~L.~Glashow,
Print-84-0577 (BOSTON);
G.~Lazarides, C.~Panagiotakopoulos and Q.~Shafi,
Phys.\ Lett.\ B {\bf 315}, 325 (1993)
[Erratum-ibid.\ B {\bf 317}, 661 (1993)]
[arXiv:hep-ph/9306332].

\bibitem{SM2}
S.~McReynolds,
``Five-dimensional Yang-Mills-Einstein Supergravity on Orbifolds: Parity assignments,"
Forthcoming.

\bibitem{DGKL:01nov}
G.~A.~Diamandis, B.~C.~Georgalas, P.~Kouroumalou and A.~B.~Lahanas,
``On the stability of the classical vacua in a minimal SU(5) 5-D  supergravity
model,''
New J.\ Phys.\  {\bf 4}, 1 (2002)
[arXiv:hep-th/0111046].

\bibitem{flipped}
S.~M.~Barr,
``A New Symmetry Breaking Pattern For SO(10) And Proton Decay,''
Phys.\ Lett.\ B {\bf 112}, 219 (1982);
A.~De Rujula, H.~Georgi and S.~L.~Glashow,
``Flavor Goniometry By Proton Decay,''
Phys.\ Rev.\ Lett.\  {\bf 45}, 413 (1980);
I.~Antoniadis, J.~R.~Ellis, J.~S.~Hagelin and D.~V.~Nanopoulos,
``Supersymmetric Flipped SU(5) Revitalized,''
Phys.\ Lett.\ B {\bf 194}, 231 (1987).

\bibitem{CGP:04jan}
M.~Cvetic, G.~W.~Gibbons and C.~N.~Pope,
``Ghost-free de Sitter supergravities as consistent reductions of string and
M-theory,''
arXiv:hep-th/0401151.

\bibitem{ES:96nov}
E.~R.~Sharpe,
``Boundary superpotentials,''
Nucl.\ Phys.\ B {\bf 523}, 211 (1998)
[arXiv:hep-th/9611196].

\bibitem{gauged}
M.~Gunaydin, G.~Sierra and P.~K.~Townsend,
Phys.\ Lett.\ B {\bf 144}, 41 (1984).

\bibitem{GZ:00}
M.~Gunaydin and M.~Zagermann,
Phys.\ Rev.\ D {\bf 62}, 044028 (2000)
[arXiv:hep-th/0002228];
M.~Gunaydin and M.~Zagermann,
Phys.\ Rev.\ D {\bf 63}, 064023 (2001)
[arXiv:hep-th/0004117].

\bibitem{Bergshoeff}
E.~Bergshoeff, S.~Cucu, T.~de Wit, J.~Gheerardyn, S.~Vandoren and A.~Van Proeyen,
``N = 2 supergravity in five dimensions revisited,''
Class.\ Quant.\ Grav.\  {\bf 21}, 3015 (2004)
[arXiv:hep-th/0403045].









\end{thebibliography}
\end{document}